\documentclass[prd,aps,oamsmath,amssymb,nofootinbib,preprintnumbers]{revtex4}
\pdfoutput=1
\usepackage{graphicx}% Include figure files
\usepackage{dcolumn}% Align table columns on decimal point
\usepackage{bm}% bold math
\usepackage{amsmath}
\usepackage{amsthm}

\usepackage{amsfonts}
\usepackage{euscript,bbm}
\usepackage{ifthen}
\usepackage{psfrag}
\usepackage{slashed}
\usepackage{hyperref}
\usepackage{textcomp,gensymb}
\def\ls{\mathrel{\lower4pt\vbox{\lineskip=0pt\baselineskip=0pt
           \hbox{$<$}\hbox{$\sim$}}}}
\def\gs{\mathrel{\lower4pt\vbox{\lineskip=0pt\baselineskip=0pt
           \hbox{$>$}\hbox{$\sim$}}}}
\def\drawbox#1#2{\hrule height#2pt

\hbox{\vrule width#2pt height#1pt \kern#1pt
              \vrule width#2pt}
              \hrule height#2pt}

\def\Asym#1#2{\vcenter{\vbox{\drawbox{#1}{#2}
              \kern-#2pt       % line up boxes
              \drawbox{#1}{#2}}}}

\newcommand{\uI}{\ensuremath{{}^{235}{\rm U}}}
\newcommand{\uII}{\ensuremath{{}^{238}{\rm U}}}
\newcommand{\uIII}{\ensuremath{{}^{239}{\rm U}}}
\newcommand{\puI}{\ensuremath{{}^{239}{\rm Pu}}}
\newcommand{\puII}{\ensuremath{{}^{241}{\rm Pu}}}
\newcommand{\np}{\ensuremath{{}^{239}{\rm Np}}}
\newcommand{\ger}{\ensuremath{{}^{72}{\rm Ge}}}
\newcommand{\si}{\ensuremath{{}^{28}{\rm Si}}}
\newcommand{\cns}{CE\ensuremath{\nu}NS}

\newcommand{\be}{\begin{equation}}
\newcommand{\ee}{\end{equation}}
\newcommand{\bea}{\begin{eqnarray}}
\newcommand{\eea}{\end{eqnarray}}

\newcommand{\gsim}{\lower.7ex\hbox{$\;\stackrel{\textstyle>}{\sim}\;$}}
\newcommand{\lsim}{\lower.7ex\hbox{$\;\stackrel{\textstyle<}{\sim}\;$}}

\begin{document}

\begin{flushright}
MI-TH-1533
\end{flushright}

\title{Sensitivity to $Z$-prime and non-standard neutrino interactions from\\ultra-low threshold neutrino-nucleus coherent scattering}

\author{Bhaskar Dutta$^{1}$}
\author{Rupak Mahapatra$^{1}$}
\author{Louis E. Strigari$^{1}$}
\author{Joel W. Walker$^{2}$}
\affiliation{$^{1}$~Mitchell Institute for Fundamental Physics and Astronomy, \\
Department of Physics and Astronomy, Texas A\&M University, College Station, TX 77843-4242, USA
}
\affiliation{$^2$~Department of Physics, Sam Houston State University, Huntsville, TX 77341, USA}

\begin{abstract}
We discuss prospects for probing $Z$-prime and non-standard neutrino interactions using 
neutrino-nucleus coherent scattering with ultra-low energy ($\sim 10$~eV) threshold Si and Ge detectors.  
The analysis is performed in the context of a specific and contemporary reactor-based experimental
proposal, developed in cooperation with the Nuclear Science Center at Texas A\&M University,
and referencing available technology based upon economical and scalable detector arrays.
For expected exposures, we show that sensitivity to the $Z$-prime mass is on the
order of several TeV, and is complementary to the LHC search with low mass detectors in the near term.
This technology is also shown to provide sensitivity to the neutrino magnetic moment, at a level
that surpasses terrestrial limits, and is competitive with more stringent astrophysical bounds. 
We demonstrate the benefits of combining silicon and germanium detectors
for distinguishing between classes of models of new physics, and for
suppressing correlated systematic uncertainties.
\end{abstract}
%MIFPA-13-15

\maketitle

%%%%%%%%%%%%%%%%%%%%%%%%

\section{Introduction} 
Coherent elastic neutrino-nucleus scattering (\cns) is a long-standing prediction of the
Standard Model~\cite{Freedman:1973yd}, and has been proposed as a new channel
to probe neutrino physics and astrophysics~\cite{Freedman:1977xn,Cabrera:1984rr}.
More recently, this process has been identified as the ultimate background to future direct dark matter
detection experiments due to neutrinos from the Sun, atmosphere, and supernovae~\cite{Monroe:2007xp,Strigari:2009bq,Billard:2013qya}. 
Because of this sensitivity to \cns{} from astrophysically-produced neutrinos,
future dark matter detectors may provide a means to probe exotic neutrino 
properties and interactions~\cite{Pospelov:2011ha,Harnik:2012ni}.

In addition to detectors developed for dark matter searches, several other source and detector configurations have been considered
to study \cns{}~\cite{Barranco:2007tz,Adams:2015ogl,Kosmas:2015sqa}.
These include neutrinos from nuclear reactors, intense radioactive sources, or from accelerators. 
However in spite of the large cross section from \cns{}, enhanced approximately by the square of the number of
neutrons in the nucleus, and the sustained experimental effort, \cns{} has yet to be detected.
This is primarily because detector technology has been unable to provide
sufficiently low threshold sensitivity to register deposition of the kinetic energy
of the heavy recoiling nucleus.

In this paper, we discuss the prospects for constraining beyond-the-standard-model (BSM)
physics with neutrinos produced from nuclear reactors using new ultra-low threshold ($\sim 10$~eV) detectors.
We investigate different $Z^\prime$ models and compare the reach with the LHC.
We probe generic neutrino non-standard interaction (NSI) vertices and explore the reach for the magnetic moment of neutrino.   
We investigate sensitivity both with and without systematic errors.
We study the benefit of combining silicon and germanium detectors, which helps to distinguish between models due to the
differential coupling to neutrons and protons.  This approach also presents substantial benefits for 
bypassing the systematic error wall.

Our experimental motivation is direct and imminent, referencing in-hand technology based upon
economical and modularly scalable germanium and silicon detector arrays,
as elaborated in a parallel publication~\cite{Mirabolfathi:2015pha}.  Experimental programs that have discussed the prospects 
for detecting \cns{} from nuclear reactors~\cite{Wong:2003ht} typically reference nuclear recoil thresholds at the keV scale or 
greater, and detectors placed $\sim 30$~m from the reactor core. In contrast, the experimental program that provides motivation
for this analysis will be characterized by eV scale nuclear recoil thresholds and closer proximity ($\sim 1$~m) 
to the reactor core.  We show that these characteristics are expected to enable the first detection of \cns{},
and also facilitate probes of BSM physics.

This paper is organized as follows.
In Section~\ref{sec:reactor}, we review properties of the proposed nuclear reactor site
and contextualize the experiment in terms of other efforts toward the detection of \cns{}.
In Section~\ref{sec:scattering}, we review the physics of neutrino-nucleus coherent scattering.
In Section~\ref{sec:zprime}, we discuss sensitivity to $Z^\prime$-mediated interactions.
In Section~\ref{sec:nsi}, we discuss sensitivity to non-standard neutrino interactions.
In Section~\ref{sec:diff}, we explore the benefits of comparing scattering rates from different nuclei to the suppression of systematic errors. 
In Section~\ref{sec:conclusions}, we present our conclusions.

\section{Reactor properties and experimental context}
\label{sec:reactor} 
The current analysis is performed in the context of a specific and contemporary experimental
proposal, developed in cooperation with the Nuclear Science Center at Texas A\&M University (TAMU),
which administrates a megawatt-class TRIGA-type pool reactor stocked with low-enriched ($\sim 20\%$) \uI{}.
Tight physical proximity of the experimental apparatus to the reactor core
can be maintained, and we will reference a baseline separation of 1~m;
the installed distance to core is expected to be within the range of 1-3~m.
This adjacency geometrically enhances the neutrino flux
to a level order-comparable with that typifying experiments
at a 30~m baseline from a gigawatt-class power reactor source. 
Reactor operators are able to provide high-precision measurements of
the thermal output power, as well as estimates (based on simulation with
the code {\tt MCNP}~\cite{mcnp}) of the isotopic fuel composition and fission fractions  $f_i/F$,
where $f_i$ is the absolute fission rate of species $i$ and $F\equiv\sum f_i$.
Neutrino data will be collected in both on and off reactor modes, and with variations
in the position of the (rail-mounted) reactor core, in order to facilitate precise
estimates of the background rate.  The concurrent observation of residual gamma and neutron events will
provide an independent handle for estimating the underlying neutrino flux.

Thermal power is generated in the reactor by fission, via neutron capture,
of the nuclei \uI{}, \uII{}, \puI, and \puII{}.  The common isotope \uII{} (with a half-life
comparable to the age of the Earth) is not readily fissile by thermal neutrons,
but will split upon fast neutron capture.  Capture of thermal neutrons will typically
induce breeding to \uIII{}, which proceeds in two steps by $\beta$-decay (with a half life
of 23 minutes and 2.4 days, respectively) to \np{} and then \puI{}.  This breeding process
also contributes subdominantly to the reactor thermal power, and appreciably to the anti-neutrino flux.
In addition to its own fission process, \puI{} will similarly exhibit breeding,
by double neutron capture, to \puII{}.  The $\alpha$-decay of \puI{} back to
\uI{} is comparatively slow, with a half life around 24,000 years.

The TAMU reactor thermal power $W_{\rm Th} = 1.00 \pm 0.02$~MW (established by
thermodynamic balance) may be combined with the relevant fission fractions
in order to establish the intrinsic anti-neutrino flux.  The thermal energy $E_{\rm Th}$ (not
counting escaping neutrinos, but incorporating recapture of neutrons not active
in down-stream fission events) released per fission (on the order of 200~MeV)
for the primary reactor constituents~\cite{Kopeikin:2004cn} are presented in Table~\ref{tab:eperfiss}.
Since the thermal output is integrated over the cascaded decay of all
sequential daughter products, some of which are relatively long-lived, these rates are
likewise mildly dependent upon the reactor fuel evolution, and typical mid-cycle values are tabulated.
Also included in Table~\ref{tab:eperfiss} are the mean cumulative energy $\langle E_\nu \rangle$ per fission
delivered to neutrinos~\cite{Kopeikin:2004cn}, the mean number $\langle N_\nu \rangle$ of
neutrinos sharing that energy budget in the decay cascade~\cite{Vogel:1980bk}, and typical fission fractions $f_i/F$
of the TAMU research reactor (cf. Ref.~\cite{Kopeikin:2012zz} for the rates $0.56:0.07:0.31:0.06$ in a representative
power reactor, where the concentration and fission of \puI{} is much more substantial).
The average neutrino fission product is boosted to about 1.5~MeV, although energies of 10~MeV and beyond are possible.
An effective fission fraction of about 0.16 (in comparison to about 0.6 in a typical power reactor context~\cite{Kopeikin:2012zz})
may be attributed to the non-fission $\uII \to\! \puI$ breeding process; this term is not included in the normalization $F$.

\bgroup
\def\arraystretch{1.5}%
\setlength\tabcolsep{4pt}
\begin{table}[h!]
  \begin{center}
    \caption{Thermal energy released per fission, average cumulative neutrino energy per fission,
	average count of cascaded neutrino emissions per fission, and typical fission fraction
	in the TAMU research reactor are provided for the primary fissile fuel components (along
	with effective values for the non-fission uranium to plutonium breeding process).
	Dimensionful quantities are reported in MeV.}
    \label{tab:eperfiss}
    \begin{tabular}{c c c c c}
      \hline
      Nucleus & $E_{\rm Th}$ & $\langle E_\nu \rangle$ & $\langle N_\nu \rangle$ & $f_i/F$ \\
      \hline
      \uI & $201.92 \pm 0.46$ & $9.07 \pm 0.32$ & $6.14$ & $0.967$ \\
      \uII &$ 205.52 \pm 0.96$ & $11.00 \pm 0.80$ & $7.08$ & $0.013$ \\
      \puI & $209.99 \pm 0.60$ & $7.22 \pm 0.27$ & $5.58$ & $0.020$ \\
      \puII & $213.60 \pm 0.65$ & $8.71 \pm 0.30$ & $6.42$ & $<0.001$ \\
      \hline
      $\uII \to\! \puI$ & $1.95$ & $1.2$ & $2.0$ & $0.16$ \\
      \hline
    \end{tabular}
  \end{center}
\end{table}
\egroup

\begin{figure}[ht]
\centering
\hspace{-5pt}
\includegraphics[width=5.0in]{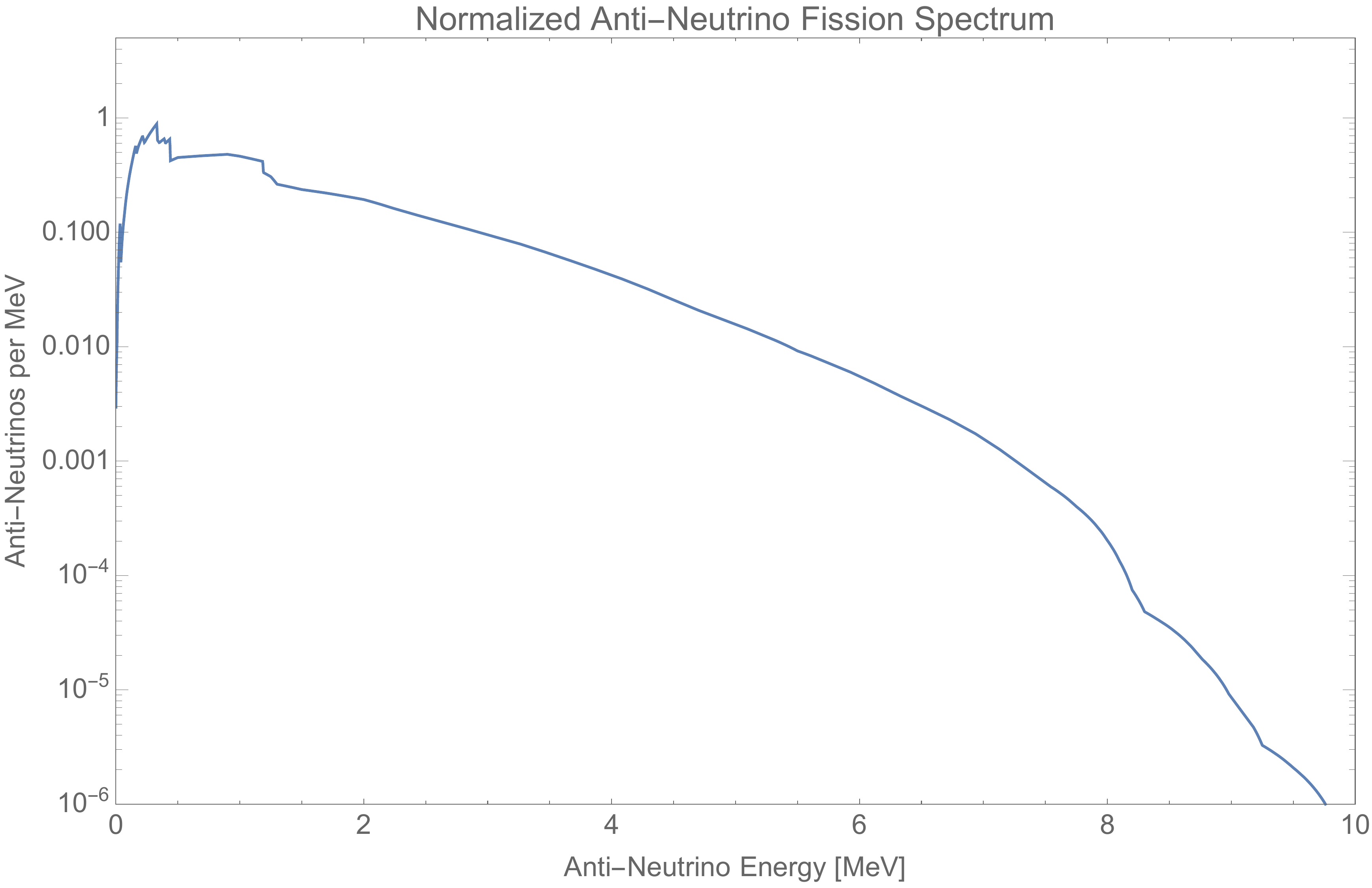}
\caption{\footnotesize Normalized anti-neutrino fission
spectrum~\cite{Schreckenbach:1985ep,Kopeikin:2012zz} employed
in the present study.}
\label{fig:spectrum}
\end{figure}

Averaging over fission fractions (which are presently dominated by \uI{}), and momentarily neglecting the
breeding of \uII{}, a thermal energy per fission of about 202~MeV is expected, with 6.1 anti-neutrinos
carrying a net invisible energy of 9.1~MeV.  At a reactor power of 1.0~MW, the extrapolated
fission rate is $3.1\times10^{16}~{\rm s}^{-1}$ with an intrinsic anti-neutrino production rate
of $1.9\times10^{17}~{\rm s}^{-1}$, and a fission anti-neutrino
flux of $\Phi_\nu^{\rm f} = 1.5\times10^{12}~{\rm cm}^{-2}{\rm s}^{-1}$
at a mean distance from core of 1~m.  If \uII{} breeding carries an effective fission
fraction of 0.16, with 2 anti-neutrino emissions per neutron capture, then there is
a second contribution to the flux of $\Phi_\nu^{\rm b} = 7.9\times10^{10}~{\rm cm}^{-2}{\rm s}^{-1}$
with a relative strength of about 5\%, but with a much more focused presentation in the low-energy regime.
These fluxes are reduced by a factor of 4 with the $1/r^2$ dilution at a distance from core of 2~m,
or a factor of 9 at 3~m.  There is additionally a subdominant solar neutrino component.

In the uranium to plutonium breeding reaction, an energy of about 2 MeV is released (about
one percent of that from a typical fission), with just under two-thirds associated with
the first (uranium to neptunium) $\beta$-transition.  The shape of the anti-neutrino
energy spectrum is modeled by the Fermi theory (cf. Ref.~\cite{0305-4616-11-3-014}
and references therein), with a functional form $E_\nu^2(Q-T_e)^2$ (where $Q$ is the energy yield,
$T_e$ the electron kinetic energy, and $E_\nu$ the neutrino energy) modified by a term representing
the Coulomb drag of the nucleus potential.  The average anti-neutrino energy budget for
the two decays (as a sum) is about 1.2~MeV, which gives each of the two decays less than half
the energy typical of a fission process.  Additionally, the normalized energy spectrum declines
precipitously above about 1.5~MeV, lacking the long tail feature exhibited by the fission products.
This imparts a distinct low-energy edge characteristic to the unified anti-neutrino spectrum,
which may be probed by an appropriately low-threshold detector.
In general, the absence of experimental data below $E_\nu = 2$~MeV is, in itself,
a motivation for direct measurements of the neutrino nuclear recoil at extraordinarily low activation thresholds.

A classic series of reactor experiments and
analyses~\cite{Vogel:1980bk,VonFeilitzsch:1982jw,Schreckenbach:1985ep,Hahn:1989zr}
performed during the 1980's are the standing primary references for numbers
on the normalization and detailed spectral distribution of neutrino daughter products;
a more recent reinterpretation of the historical data was presented in Ref.~\cite{Mueller:2011nm}.
To be concrete, all modeling in the present study will adopt standardized tabulated spectra.
Below 2~MeV (the inverse $\beta$-decay $\overline{\nu}_e+p \to e^+ + n$ threshold is $E_\nu > 1.8$~MeV)
there is no experimental data, and we will employ the theoretically established curve of
Ref.~\cite{Kopeikin:2012zz}, which assumes typical power reactor distributions for each of the
Table~\ref{tab:eperfiss} processes.  Above 2~MeV, we will employ the spectrum of Ref.~\cite{Schreckenbach:1985ep},
which is extrapolated from direct observation of positron emission in \uI{}.
Experiment and theory agree on the integrated flux at better than 2\%
above the inverse $\beta$-decay threshold~\cite{Li:2001ha,Kopeikin:2012zz}.
The resulting unified spectrum is presented graphically as Fig.~(\ref{fig:spectrum}).

There are several other existing experimental proposals (cf. Ref~\cite{McLaughlin:2015xfa})
for measurement of the \cns{} process, which employ various modes of neutrino production.
We summarize here a few relevant examples, emphasizing similarities
and differences as well as advantages and disadvantages of the various approaches.
The COHERENT~\cite{Akimov:2015nza} project is designed around the use
of a stopped-pion neutrinos derived from the Spallation Neutron Source
at Oak Ridge National Laboratory.  This production mode generates prompt
muon neutrinos, as well as delayed electron neutrinos and muon anti-neutrinos,
in contrast to the pure electron anti-neutrino content expected at a reactor source.
This fact implies that distinct interaction vertices may be probed in a complementary manner.
Also, the pulsed timing of the beam yields substantial advantages for the
reduction of asynchronous backgrounds. Moreover, the typical neutrino energy is
above 30~MeV, which is a factor of twenty or more larger than the
mean energy at a reactor.  Since the integrated cross section
scales as $E_\nu^2$ both the event rate at fixed flux and
the minimal recoil threshold are simultaneously elevated by a factor in the
neighborhood of 500.  In particular, this means that less exotic detectors,
including non cryogenic scintillators, may be considered. However, the expected
neutrino flux is substantially lower than at a reactor, on the order of
a few times $10^{7} /{\rm cm}^2/{\rm s}$ at a range of 20 [m] from the target.
In particular, this is down by about five orders of magnitude from the expected
output of the TAMU reactor at one meter.  The advent of suitably low-threshold
detectors thereby tilts the advantage in expected \cns{} event rate per
kilogram back toward reactor-based sources by as much as a few hundred-fold.
There are also proposals for accelerator-based stopped pion sources, such
as the DAE$\delta$ALUS~\cite{Alonso:2010fs} experiment, which have essential
similar characteristics.  The proposed liquid argon detector in that experiment
is projected to register approximately 10 \cns{} events per kilogram per year,
a figure that is suppressed by up to three magnitude orders relative to the
projections in the present work.  Candidates for reactor-based \cns{}
observation include the TEXONO~\cite{Singh:2003ep,Wong:2005vg}
and CoGeNT at SONGS projects.  All such reactor based environments feature an essentially
identical electron antineutrino spectrum.  Potentially distinguishing features including
the reactor power, the distance from core, and the recoil threshold sensitivity.
The TEXONO experiment is housed at the Kuo-Sheng power reactor in Taiwan,
which operates in the typical few gigawatt power range.  At thirty meters from
core, it yields a flux that is broadly comparable to (or perhaps larger by a few times than)
that available at our referenced megawatt research reactor at one meter from core.
The SONGS facility also employs a research reactor, albeit one generating approximately
30 megawatts of power.  With the detector placed 20 meters from core, this corresponds
to a net reduction of one magnitude order in flux relative to our proposal at one meter
from core, or flux parity at three meters from core.
Both projects are likewise actively pursuing low noise germanium detection environments
capable of reaching recoil thresholds in the one-to-a-few hundreds of eV
range~\cite{Soma:2014zgm,Barbeau:2007qi}.  Our referenced \ger{} and \si{}
detector technology is capable of substantially broaching the 100~eV threshold,
plausibly reaching as low as 10~eV, in the very near term future~\cite{Mirabolfathi:2015pha}.

\section{Neutrino-Atom Scattering}
\label{sec:scattering} 
The standard model (SM) electroweak Lagrangian exhibits the gauge symmetry $SU(2)_L \times U(1)_Y$,
with gauge fields $W^i_\mu$ and $B_\mu$, and couplings $g$ and $g'$,
respectively.  After electroweak symmetry breaking, the residual
symmetry is $U(1)_{EM}$, with coupling $e \equiv g \sin \theta_W$, where
the Weinberg angle is defined by the relation $\tan \theta_W \equiv g'/g$.
The massive neutral current is mediated by the mixed bosonic state $Z_\mu$,
with an effective coupling (conventionally defined as) $g/\cos \theta_W$
and an associated charge operator $T_3-Q_{EM} \sin^2 \theta_W$, where
$T_3$ is the diagonal generator of $SU(2)_L$, with eigenvalues $\pm 1/2$
for the upper and lower components of a field doublet, respectively.

For small momentum exchange, the $Z_\mu$ propagator will be dominated by its $M_Z^2$
mass-denominator, generating an effective (mass-suppressed)
dimension-six coupling $G_F \equiv \sqrt{2}g^2/8M_W^2$, where $M_W \equiv M_Z \cos \theta_W$.
The differential cross-section~\cite{Vogel:1989iv} for SM scattering of a neutrino with energy $E_\nu$
from a target particle of mass $M$ and kinetic recoil $T_{\rm R}$ is, in terms of
the applicable vector $q_V \equiv q_L+q_R$ and axial $q_A \equiv q_L-q_R$ charges, 
\be
\frac{d \sigma}{ d T_{\rm R}} = \frac{G_F^2 M}{2\pi}\bigg[\, (q_V+q_A)^2
+ (q_V-q_A)^2\left( 1-\frac{T_{\rm R}}{E_\nu} \right)^2 - (q_V^2-q_A^2)\frac{M T_{\rm R}}{E_\nu^2} \bigg]. 
\label{eq:dsdtde}
\ee
Eq.~(\ref{eq:dsdtde}) is applicable both to scattering from electrons and to scattering from nuclei.
The SM neutrino is purely left-handed, but couples via the $Z$-boson neutral current
(with distinct strengths) to both left- and right-handed fermionic currents.
By convention, the global contribution (a factor of $1/2$) to the neutral-current vector and axial charges
from the pure-left neutrino has been factored out, such that the referenced charges (and $L/R$ chirality designations)
refer only to the scattering host.

For ($e,\nu_e$) scattering, the flavor-diagonal $t$-channel exchange of a $W$-boson interferes with the neutral current term, effectively boosting
($q_V,q_A$) by a unit value~\cite{Vogel:1989iv}.  This diagram is not applicable, though, to the scenario of a reactor anti-neutrino
source.  However, there is a second subtlety that is relevant in this case, namely the emergence of a relative negative phase
between $(q_V,q_A)$ associated with the parity-flip.  Absorbing this sign into the axial coupling, the SM expressions for
anti-neutrino scattering from a generic fundamental particle target become $(q_V,q_A) \equiv (T_3-2Q\sin^2\theta_W,-T_3)$.
For coherent nuclear scattering, these terms should be summed over the quark content of protons and neutrons, and either
multiplied by the respective counts $(Z,N)$ of each (in the vector case) or multiplied by the respective differential
counts $(Z^+-Z^-,N^+-N^-)$ of up and down spins (in the axial case)~\cite{Barranco:2005yy}.
The leading event contribution comes from the neutron count $N$, with $q_V^N = -0.5$ (independent of the Weinberg angle),
whereas coupling to the proton $q_V^P \simeq +0.038$ experiences strong interference and
is relatively suppressed by more than a magnitude order.

The described sum over nuclear constituents at the coupling level, prior to squaring in the amplitude, is the essence of
the nuclear coherency boost.  By contrast, electron scattering sums over the atomic number $Z$ incoherently,
boosting the cross-section linearly rather than quadratically.  We note as a curiosity that further reduction of the
neutrino energy (to around the milli-eV scale) induces coherency at the level of electron scattering~\cite{Sehgal:1986gn}.
Both of the suggested target nuclei, namely \ger{} and \si{}, have a total spin of zero,
although the germanium nucleus has a deficit of spin-up protons (and an excess of spin-up neutrons)
of two units, which is observed to boost the expected scattering count by about 3.5\% relative to the dominant vector mode.
There are calculable correction factors that account for the eV-scale binding energy of scattered electrons,
which we neglect in comparison to the MeV-scale energy of the reactor anti-neutrino source.
Another subleading effect that that could contribute meaningful systematics on the order of targeted
signals of new physics is the excitation of low-lying nuclear states.
We generally expect charged current, incoherent, and inelastic scattering cross sections
to increase more rapidly with energy than the \cns{} cross section~\cite{Brice:2013fwa},
such that these effects are substantially suppressed for reactor-scale neutrino energies in the few MeV range.
In keeping, we find the calculable form factors~\cite{Engel:1991wq} that gauge applicability of the nuclear
coherency assumption to be of unit value within a part per few thousand at typical reactor neutrino energies,
and consequently further neglect their consideration here.
Incidentally, such effects may be more problematic (or one may have  greater opportunity to probe
structure of the nuclear form factor) with a higher energy neutrino source, e.g. a stopped pion beam.
Radiative corrections to the sine-squared Weinberg angle and the neutral current
couplings~\cite{Eidelman:2004wy} have been included in leading terms.

\begin{figure}[ht]
\centering
\hspace{-5pt}
\includegraphics[width=3.2in]{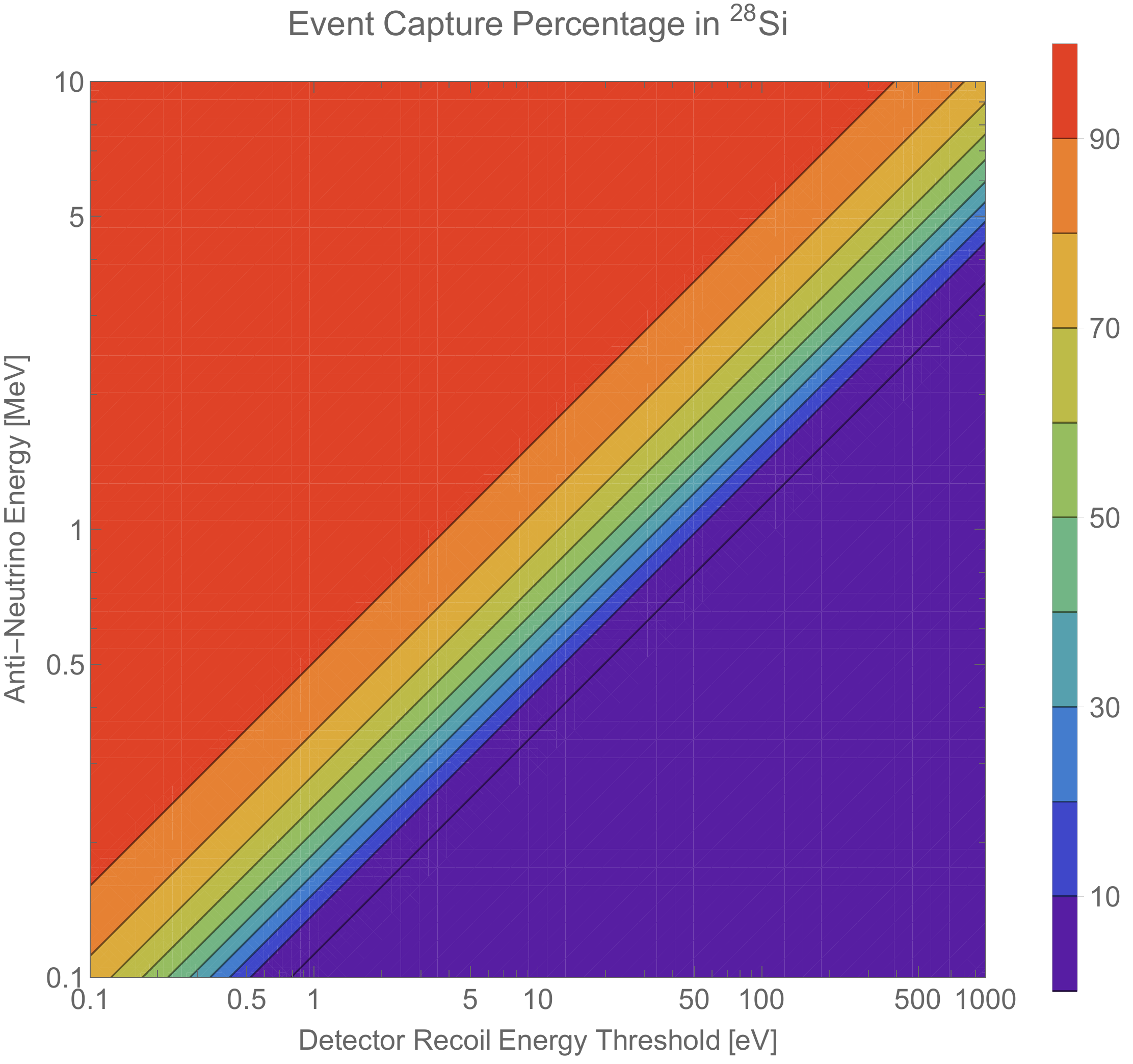}
\includegraphics[width=3.2in]{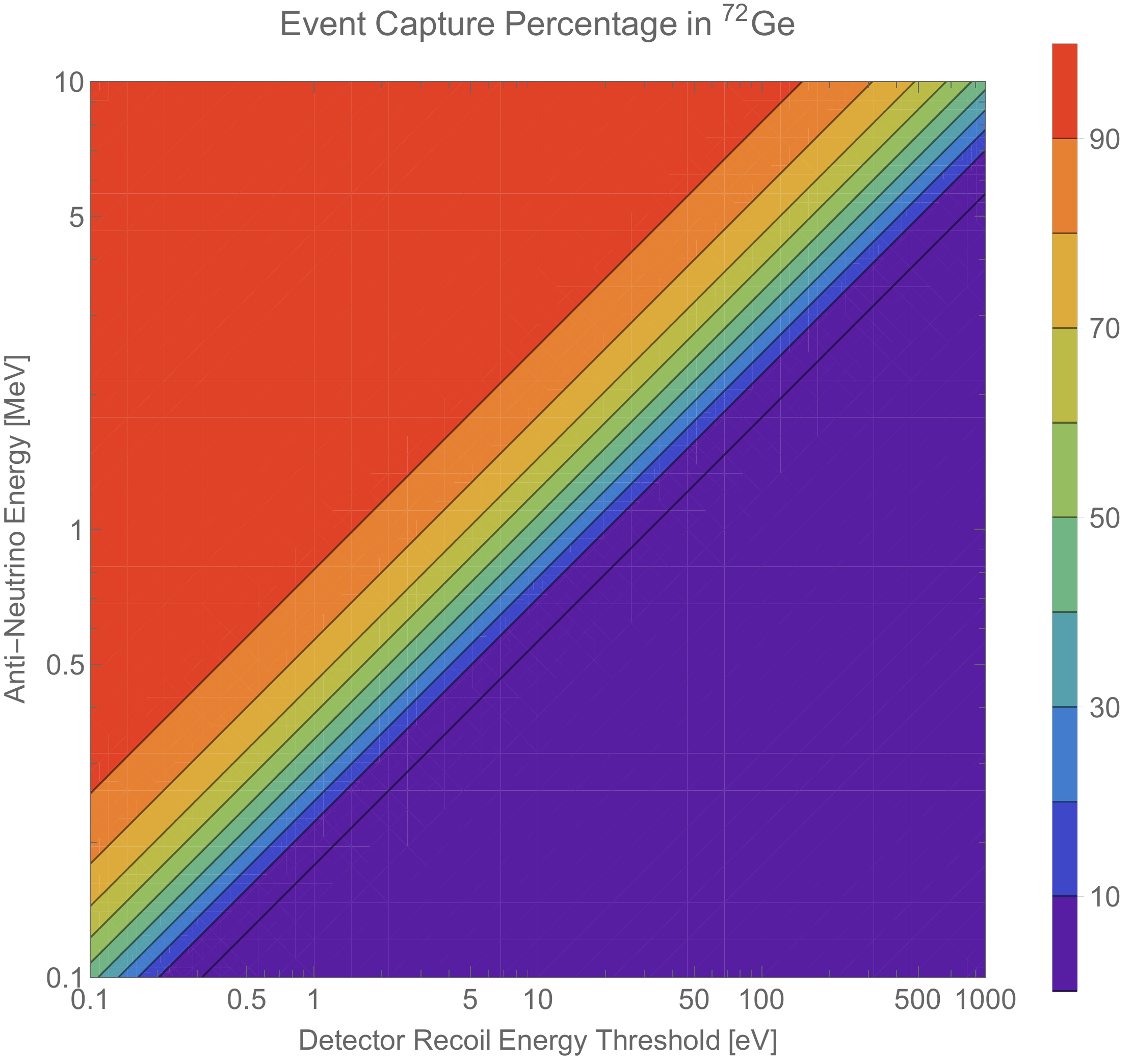}
\caption{ \footnotesize Percentage of nuclear recoils captured by \si{} and \ger{} detectors
as a function of the incident anti-neutrino energy
$E_\nu$ and the recoil threshold $T_{\rm R}^{\rm th}$.}
\label{fig:capture}
\end{figure}

It is clear that nuclear scattering will generally be dominated by the vector charge, and in the limit of
vanishing axial charge the residual functional dependence $1-M T_{\rm R}/2E_\nu^2$ interpolates between
a large cross-section at zero recoil and a vanishing cross-section at cut-off, where energy-momentum
conservation stipulates the maximum recoil $T_{\rm R}^{\rm max} = 2E_\nu^2/(M+2E_\nu)$ achievable
in a collision with no glancing component.  The large mass-denominator in this term highlights the
necessity of ultra-low threshold detectors for observation of the heavily boosted \cns{} feature.
We calculate that a detector threshold around 50, 20~eV is required in \si, \ger
in order to capture about half of the scattering from fission neutrinos
with a mean energy of 1.5~MeV, as demonstrated graphically in Figs.~(\ref{fig:capture}).
We note additionally that the area under the $1 - T_{\rm R}/T_{\rm R}^{\rm max}$ curve is
$T_{\rm R}^{\rm max}/2 \simeq E_\nu^2/M$, which yields the previously quoted scaling
with regards to the incident neutrino energy, and the associated prospect that higher
values of $E_\nu$ may partially offset very stringent recoil threshold requirements.

If the neutrino has a non-vanishing magnetic moment $\mu_\nu$ (expressed dimensionlessly
as a multiple of the Bohr magneton $\mu_{\rm Bohr} \equiv e/2m_e$), then this supplements (as a simple sum) the scattering
cross section(s) described in Eq.~(\ref{eq:dsdtde})~\cite{Vogel:1989iv}, 
\be
\frac{d \sigma}{d T_{\rm R}} \bigg\vert_{\mu_\nu} = \,\, \frac{\pi \alpha^2 \mu^2_\nu}{m^2_e}\bigg[\, 
\frac{1-T_{\rm R}/E_\nu}{T_{\rm R}} + \frac{T_{\rm R}}{4E^2_\nu} \bigg]. 
\label{eq:dsdtdeMu}
\ee
The second term of Eq.~(\ref{eq:dsdtdeMu}) applies only to the case of nuclear scattering, while both nuclear
and electron scattering reference the first term.  For coherent nuclear scattering, the unified contribution will
again be multiplied by $Z^2$, whereas the sum over individual elements of the electron cloud is again linear, providing a factor of just $Z$.
For nuclei with odd atomic number there are additional terms dependent upon the nuclear magnetic moment~\cite{Vogel:1989iv}.

\begin{figure}[ht]
\centering
\hspace{-5pt}
\includegraphics[width=5.0in]{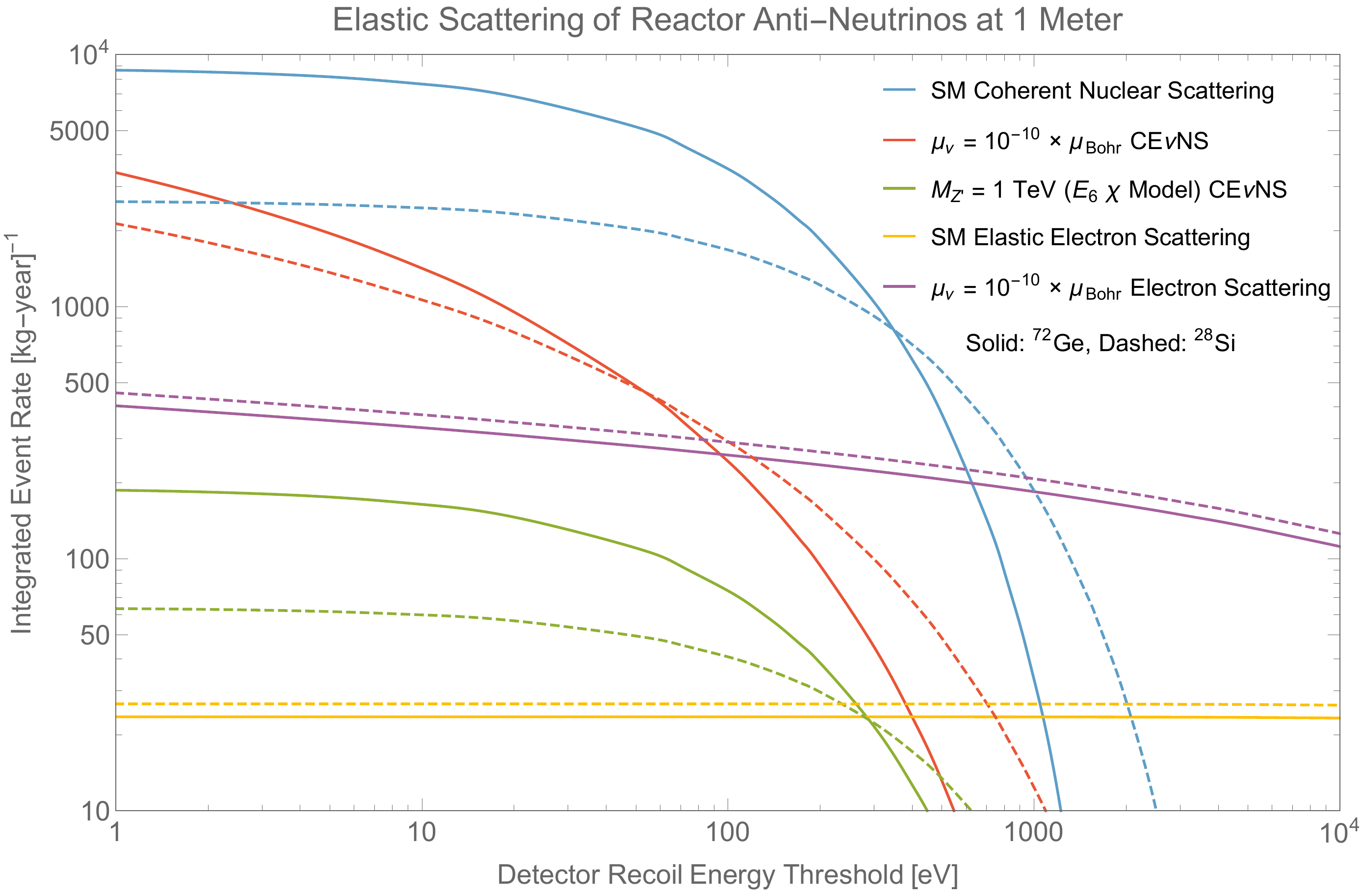}
\caption{ \footnotesize Integrated yield of nuclear recoil events
(per kg per year) captured by \ger{} (solid) and \si{} (dashed) detectors
as a function of detector recoil threshold $T_{\rm R}^{\rm th}$.
Independent curves are exhibited for the contribution of nuclear and electron cloud recoils,
and for benchmark $Z^\prime$ and neutrino magnetic moment $\mu_\nu$ scattering.
This plot cannot be used to compare visibility of magnetic moment
and $Z^\prime$ searches, as both use an arbitrarily established benchmark value.}
\label{fig:integrated_rate}
\end{figure}

In order to compute the cumulative expected Standard Model anti-neutrino capture (cf. Ref~\cite{Barranco:2005yy}),
it is necessary to integrate in the region of the $E_\nu$ vs. $T_{\rm R}$ plane that is
above $E_\nu > E_\nu^{\rm min} = (T_{\rm R}+\sqrt{2 M T_{\rm R} + T_{\rm R}^2})/2$, which is the minimal neutrino energy
(i.e. the inversion of the expression for $T_{\rm R}^{\rm max}$) required to trigger a given recoil,
and for $T_{\rm R} > T_{\rm R}^{\rm th}$ of the detector recoil threshold.
The integrand is a product of the previously described differential cross section and the
normalized anti-neutrino energy spectral distribution $dN_\nu/dE_\nu \div N_\nu$, as well as
the ambient anti-neutrino flux, the detector mass, and the exposure time.

Fig.~(\ref{fig:integrated_rate}) exhibits primary results for the expected event capture rate,
as a function of the detector recoil threshold $T_{\rm R}^{\rm th}$, for both \ger{} (solid curves)
and \si{} (dashed curves), per kilogram, per year.  Separate curves are presented for nuclear and
electron recoils, for the magnetic moment contribution (at a benchmark value of $\mu = 10^{-10}$),
and for the enhancement expected from existence of a $Z^\prime$ gauge boson
(using a benchmark $E_6$ ``$\chi$'' model with $M_{Z^\prime} = 1$~TeV),
the treatment of which are described in detail subsequently.
The electron curves all flatten out at much larger thresholds, due to the denominator $M+2E_\nu$ in the maximum recoil. 
It is clear that SM elastic scattering from the electron cloud is important at super-keV thresholds,
but is completely lost relative to the emergence of the CE$\nu$NS at low thresholds.
The magnetic moment contribution is larger in terms of absolute numbers for the nucleus, although it is
a more relevant fractionally for the electrons.
The recoil threshold $T_{\rm R}^{\rm th}$ contributes in the denominator of a logarithm
to the integrated event rate for the magnetic moment scattering,
which explains the steepness of the observed slope for the red \cns{} curves.  There is also a mass denominator inside this
logarithm that suppresses the nuclear magnetic moment scattering, but the enhancement for the electrons is
regulated by a sum with the neutrino energy, and its effect is limited.
It cannot overcome the coherent nucleon-squared enhancement that sits outside the logarithm for the nuclear case.
Electron magnetic moment scattering is comparable to nuclear magnetic
moment scattering at a threshold around 100~eV, the nuclear term dominates by a factor of about 5 around 10~eV,
and a factor of about 10 at 1~eV.  The benefit of low thresholds to the observation
of magnetic moment interactions is thus very clear, and even more so than for $Z^\prime$ scattering,
with event rates rising much faster in this regime than for all other CE$\nu$NS processes.

\section{Sensitivity to $Z$-prime Interactions}
\label{sec:zprime} 
In this section, we discuss prospects for constraining $Z^{\prime}$ interactions~\cite{Robinett:1982tq, Hewett:1988xc, Carena:2004xs, Langacker:2008yv} 
using ultra-low energy threshold Si and Ge detectors~\cite{Barranco:2007tz}.
Heavy analogs of the $Z$ boson are a mainstay of BSM physics, associated with extra local $U(1)$
symmetries that arise naturally in various string theoretic and grand unified theory (GUT) constructions.
The presence of a heavy neutral $Z^\prime$ gauge boson would manifest itself as a modification
(generally enhancement) in the rate of detected anti-neutrino scattering events.  Specifically,
the SM vector and axial charges are summed (there can be interference) with contributions
to $(q_V,q_A)$ from the new physics, which will necessarily carry the dependence $(M_Z/M_{Z^\prime})^2$,
reflecting modification of the propagator to the Fermi coupling.
The event rate is proportional to the charge-squared, and the new physics will manifest
primarily via a cross-term with the much larger SM charges, such that the expected signal
event rate declines still as just  $1/M_{Z^\prime}^2$.
Also of potential interest, although beyond the present scope, is the possibility of
light Abelian vector bosons with Stueckelberg mass generation~\cite{Laha:2013xua,Ng:2014pca}.

In order to formulate a charge factor for the new physics that sums correctly
with the SM terms, several normalization coefficients must be computed.  In full, the prescription
for a scattering from a target $i$ will be
\be
\quad Q_{SM}(i) \,\, \Rightarrow \,\, Q_{SM}(i) \,+\,  Q_{\rm BSM}(i) \,\,\times
\bigg\{ Q_{\rm BSM}(\nu) / Q_{\rm SM} (\nu) \times (g'\, \cos \theta_W/g)^2 \,\times (M_Z/M_{Z^\prime})^2 \bigg\} \nonumber \,,
\label{eq:bsmprescription}
\ee
where previously described global coupling, charge, and mass terms that were explicitly factored out in the SM
analysis have been exchanged for the appropriate BSM analogs;
$g'$ is the BSM hypercharge, and we assume a decoupling limit where the
heavy $Z^\prime$ does not mix in the electroweak symmetry breaking.

In order to broadly assess the sensitivity of a solid state detector to these heavy particles,
we consider five benchmark $Z^{\prime}$ models that are representative of the most common approaches
to this idea.  Two of the benchmarks will be taken from the symmetry breaking of a typical string-derived
scenario featuring the unified $E_{6}$ group.  In this scenario, two $U(1)$ factors may combine via
an unspecified mixing angle $\beta$ to generate a single TeV-scale gauge field.  The most studied
mixing angles are $\beta=0$ ($\chi$ model), $\beta = \cos^{-1}\sqrt{3/8}$ ($\eta$ model), and $\beta = \pi/2$ ($\psi$ model).
Silicon and germanium detectors are not sensitive to the $\psi$ model, which features purely axial couplings.
Two additional benchmarks are associated with models invoking a baryon minus lepton $B-L$
symmetry~\cite{Basso:2009gg,Basso:2010pe}.  As this is not a unified theory, the coupling strength is arbitrary, and we consider both
$(g'=0.4)$ and $(g'=0.2)$ examples, the former being near to the limit allowed by high-energy
consistency of the renormalization group.  The final model considered is a toy model called the
sequential standard model (SSM), whose couplings are identical to those of the $Z$-boson after
electroweak symmetry breaking.  

Table~\ref{tab:charges} itemizes the quark and lepton neutron current
charges in the SM (equivalently SSM), and the various described extensions~\cite{Langacker:2008yv,Basso:2009gg,Gershtein:2013iqa}.
In the $E_6$ models, $(g'/g)^2$ is fixed to be $\frac{5}{3} \tan^2\theta$,
where the factor of $5/3$ preserves proper GUT charge normalization.
In the $B-L$ models, $(g'/g)$ is not constrained, and the coupling $g'$ is generally provided explicitly;
the $SU(2)_L$ coupling is given numerically by $g \simeq 0.65$.
In the sequential standard model, $(g'\cos\theta_W/g)^2 = 1$.
Continuous parameterization of $E_6$ models is provided by the definition
$Q_{E_6} \equiv \cos\beta\,Q_{E_6}^\chi + \sin\beta\,Q_{E_6}^\psi$.
As before, the unified axial charge $q_A$ will inherit a relative sign-flip
for the scenario of anti-neutrino scattering, after summation of all relevant contributions.

\bgroup
\def\arraystretch{1.5}%
\setlength\tabcolsep{6pt}
\begin{table}[h!]
  \begin{center}
    \caption{Quark and lepton neutral current charges in the SM and various extensions.}
    \label{tab:charges}
    \begin{tabular}{c c c c c}
      \hline
       & $Q_{SM}$ & $\sqrt{40}\, Q_{E_6}^\chi$ & $\sqrt{24}\, Q_{E_6}^\psi$ & $Q_{B-L}$ \\ 
      \hline
      $u_L$ & $\frac{1}{2} - \frac{2}{3}\sin^2\theta_W$  & $-1$ & $1$  & $\frac{1}{3}$ \\
      $d_L$ & $-\frac{1}{2} + \frac{1}{3}\sin^2\theta_W$ & $-1$ & $1$  & $\frac{1}{3}$ \\
      $u_R$ & $ - \frac{2}{3}\sin^2\theta_W$             & $1$  & $-1$ & $\frac{1}{3}$ \\
      $d_R$ & $ \frac{1}{3}\sin^2\theta_W$               & $-3$ & $-1$ & $\frac{1}{3}$ \\
      $\nu_L$ & $\frac{1}{2}$                            & $3$  & $1$  & $-1$ \\
      $e_L$ & $-\frac{1}{2} + \sin^2\theta_W$            & $3$  & $1$  & $-1$ \\
      $e_R$ & $ \sin^2\theta_W$                          & $1$  & $-1$ & $-1$ \\
      \hline
    \end{tabular}
  \end{center}
\end{table}
\egroup

We note that bounds on the $Z^{\prime}$ from \cns{} are complementary to those obtained from the LHC,
which probes for resonance peaks in the dilepton invariant mass. Whereas a collider is directly
sensitive to the $Z^\prime$ mass scale, individual coherent nuclear recoil events cannot tag the
mass of the mediating species. Still, the coherency boost at very low recoil thresholds can allow
for exquisite statistical sensitivity to the $Z^\prime$ scale.
Projected 95\% confidence limits on the $Z^{\prime}$ mass at the $\sqrt{s} = 13/14$~TeV LHC with a few (1-3) hundreds
of events per femtobarn of luminosity are in the range of 5 to 6 TeV~\cite{Basso:2010pe,Gershtein:2013iqa} for
the described benchmark scenarios.  Limits for the $B-L$ model go as (6,5,4.4) TeV for couplings $g'$
of (0.4,0.2,0.1), respectively.  The naive proportional scaling of the mass limit with the $g'$ coupling is not
realizable in a collider scenario, where mass suppression of the parton luminosity at high momenta inhibit the
reach into heavier models.  The solid state detectors do not exhibit this shortcoming, being sensitive to the
$Z^{\prime}$ mass only in the off-shell Fermi-coupling sense, and thus fare proportionally better at large coupling.

\begin{figure}[htbp]
\begin{center}
\includegraphics[width=5.0 in]{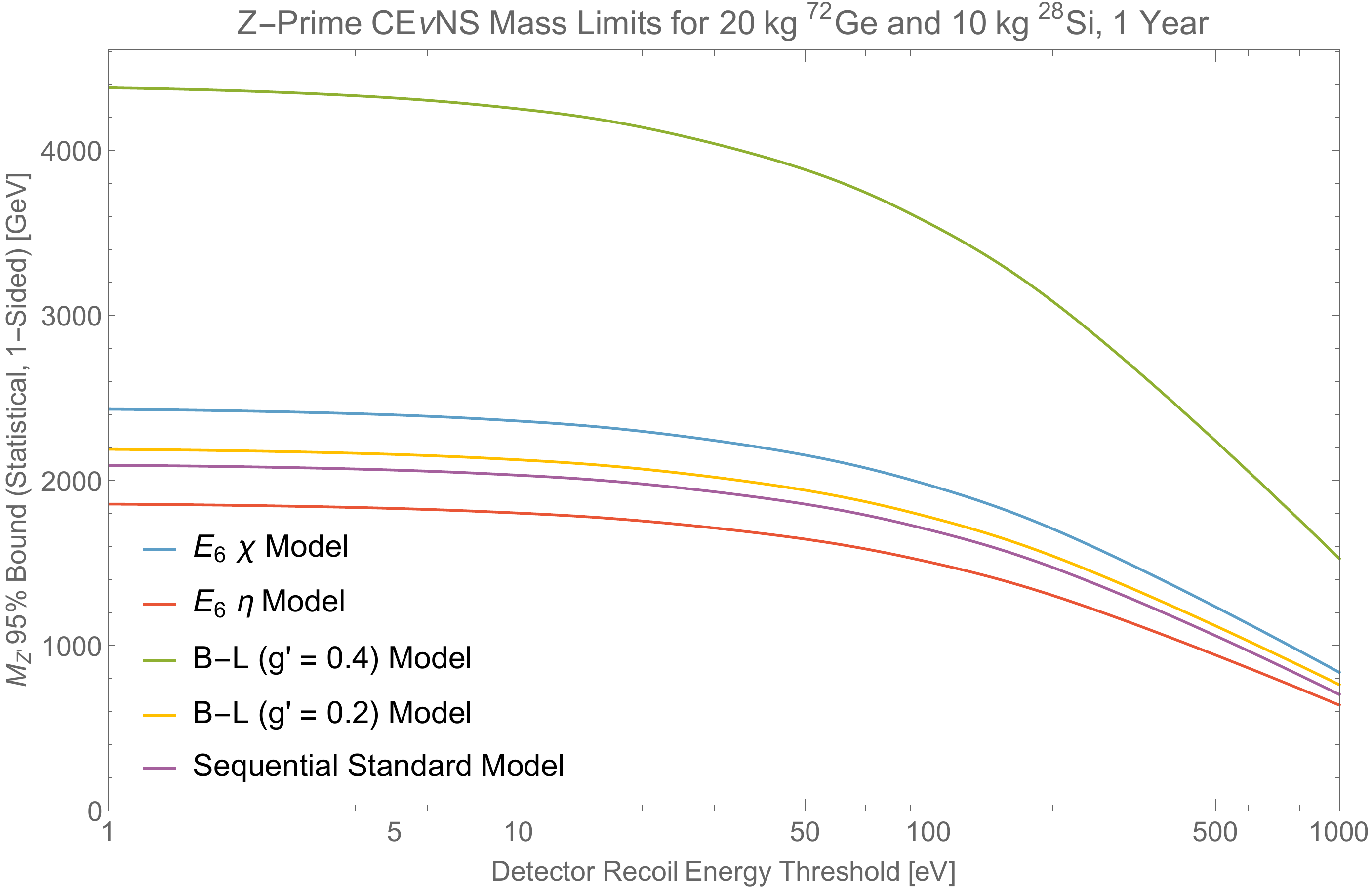}
\caption{\footnotesize Sensitivity to $Z^{\prime}$ in different models.}
\label{fig:zprime}
\end{center}
\end{figure}

In order to concretely assess event detection significance, we select the signal $S$ to background $B$ significance metric $S/\sqrt{B}$.
Both $S$ and $B$ scale linearly with luminosity $\mathcal{L}$, and $S$ additionally scales proportionally with $1/M_{Z^\prime}^2$.
At fixed significance $S/\sqrt{B}$, the mass reach for nuclear recoil detectors will thus scale like
$M_{Z^\prime}^{\rm Lim} \propto \mathcal{L}^{1/4}$ i.e. a fourth-root of the product of
neutrino flux, exposure time, and detector mass.  Projected mass exclusions at 95\% confidence (statistical only, single-sided)
for each benchmark model are provided as a function of the detector recoil threshold $T_{\rm R}^{\rm th}$ in
Fig.~(\ref{fig:zprime}).
The benchmark early phase detector is composed of \ger{} and \si{} in roughly a 2:1 mass ratio, with a combined mass of 30~kg, operating
for a one-year continuous exposure, at 1~m from core.  Bounds are in the range of 1.8 to 2.4~TeV for most models, reaching
above 4~TeV for the strongly-coupled $B-L$ model.  Expanding upon this example, 
scaling up to 5 ton-years or 100 ton-years would increase the bounds by factors of almost 4 and 8, respectively.
This suggests that a low threshold \cns{} measurement could be competitive with the foreseeable collider
reach, and even substantially exceed it, given sufficient scaling of the experiment.

However, a full analysis of the $Z^\prime$ scale bounds from \cns{} will necessarily be dominated by
statistical uncertainties. Although detection efficiency above threshold and within the fiducial volume
approaches 100\%, with controllable backgrounds, and vanishing pile-up (milli-second recovery time),
the dominant uncertainty will be propagated from errors in the reactor thermal power, and from the
extrapolation of this power into the associated anti-neutrino spectrum. Uncertainty estimates on the order
of 2\% are typical, although it may be possible to reduce this to around a half of a percent (cf. Ref.~\cite{Cao:2011gb}).

\begin{figure}[htbp]
\begin{center}
\includegraphics[width=5.0 in]{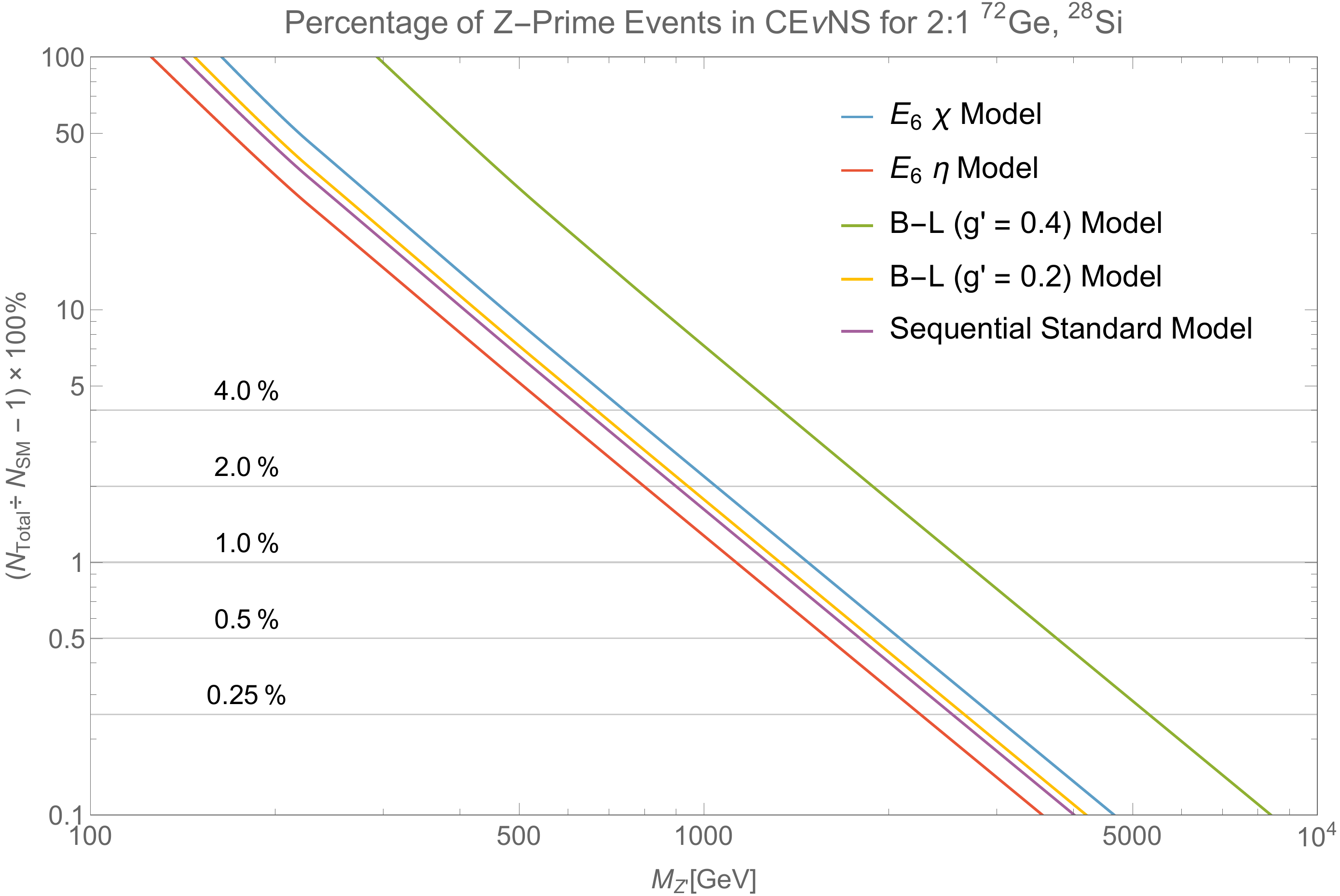}
\caption{\footnotesize $Z^{\prime}$
contributions to event rates in different models,
as a fraction of the Standard Model
$Z$-mediated background.}
\label{fig:bsmpercent}
\end{center}
\end{figure}

In Fig.~(\ref{fig:bsmpercent}), we show the BSM event fraction for various $Z^\prime$ models, as a function
of $M_{Z^\prime}$, which should not be less than the order of the anticipated systematic uncertainties.
Such fixed percentile errors will do more damage in the large detector mass and high luminosity regimes,
where statistical fluctuations are tailing off as a percentage of events.  This does not, however, imply that
additional statistical resolution is without benefit.  Since various $Z^\prime$ models couple
distinctly to up and down quarks, differential and rational event counts in detectors with
contrasting atomic and mass numbers, such as \ger{} and \si{}, can be very sensitive to
deviations from the standard model in a manner that cancels leading systematic uncertainties.
This sensitivity to the {\it existence and mode} of new $Z^\prime$ or $\mu_\nu$ physics, even more
so than the scale, is a key distinguishing benefit nuclear recoil detectors over other approaches~\cite{Barranco:2005yy},
as will be further elaborated in the final section.

We close this section by noting that ATLAS~\cite{Aad:2015owa} has recently reported excesses in searches for massive resonances decaying
into a pair of weak gauge bosons, and CMS~\cite{Khachatryan:2014hpa} also has reported a diboson excess.
One suitable explanation exists in the context of a leptophobic
$SU(2)_L\times SU(2)_R\times U(1)_{B-L}$ model~\cite{Gao:2015irw}.
The prediction arising from this model is a 3-5 TeV $Z^{\prime}$ that couples to SM
leptons as shown in Ref.~\cite{Gao:2015irw}.  This sort of field is potentially
well-suited for study via the \cns{} approach, especially with regards to the
probing its characteristic mode of coupling to up- and down-quarks.  Such complementary
approaches are very useful in establishing a particular model of new physics.
However, even in the absence of new physics, a first detection of the \cns{} process (which is of
substantial interest in its own right), and the accumulation of  additional statistical resolution,
will allow for the SM neutrino interactions to be studied in fine detail.  Such observations are
independent of the inverse $\beta$-decay detection mode, and provide access to those portions
of the neutrino spectrum that are below the kinematic threshold for this process.

\section{Sensitivity to neutrino Non-standard interactions and magnetic moment} 
\label{sec:nsi} 

\begin{figure}[htbp]
\begin{center}
\includegraphics[width=5.0 in]{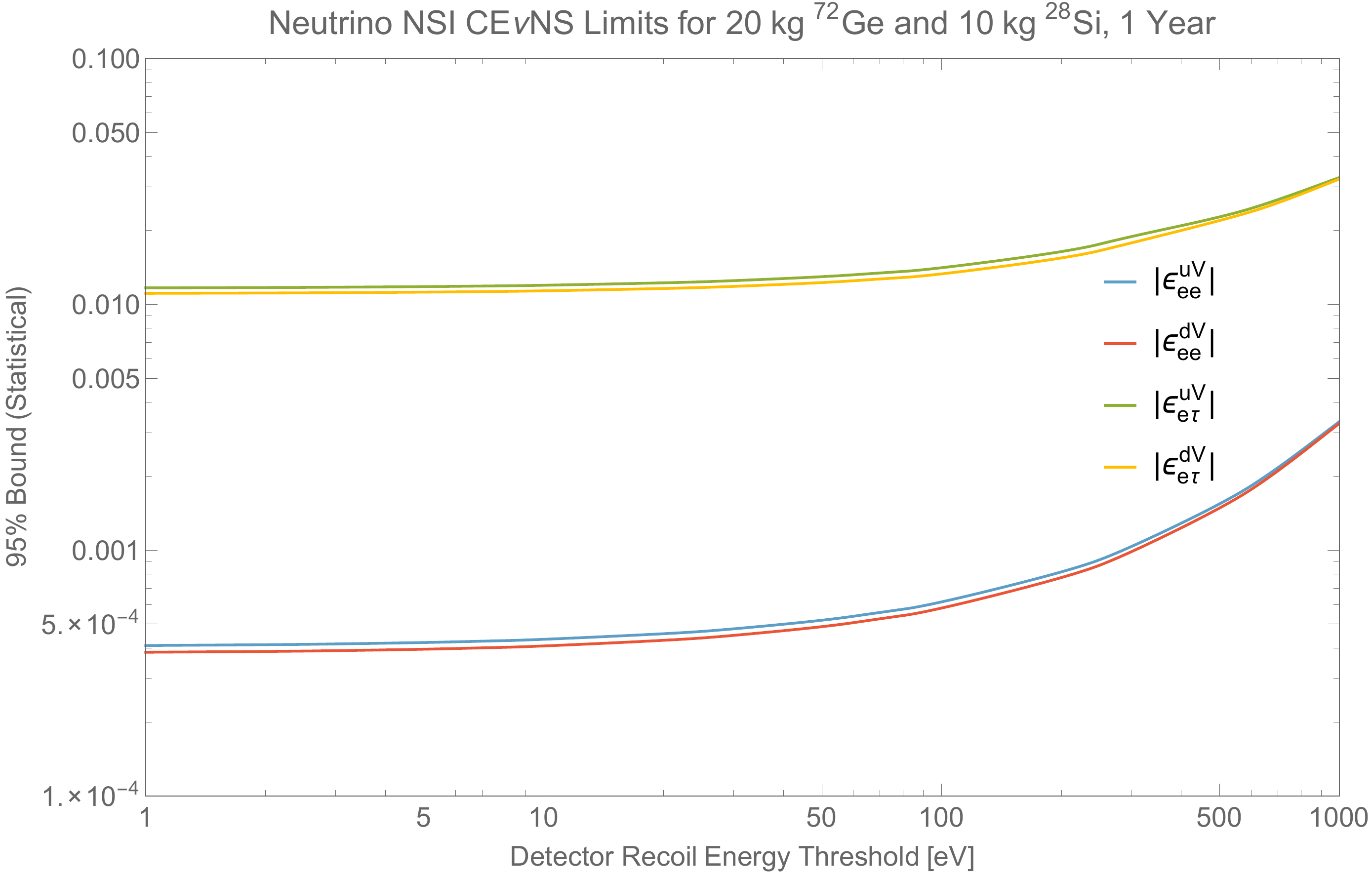}
\caption{\footnotesize
$Z^{\prime}$ Sensitivity to
non-standard neutrino interactions, including flavor-changing terms, in a low-mass
\ger{} and \si{} detector.}
\label{fig:nsi}
\end{center}
\end{figure}

\begin{figure}[htbp]
\begin{center}
\includegraphics[width=5.0 in]{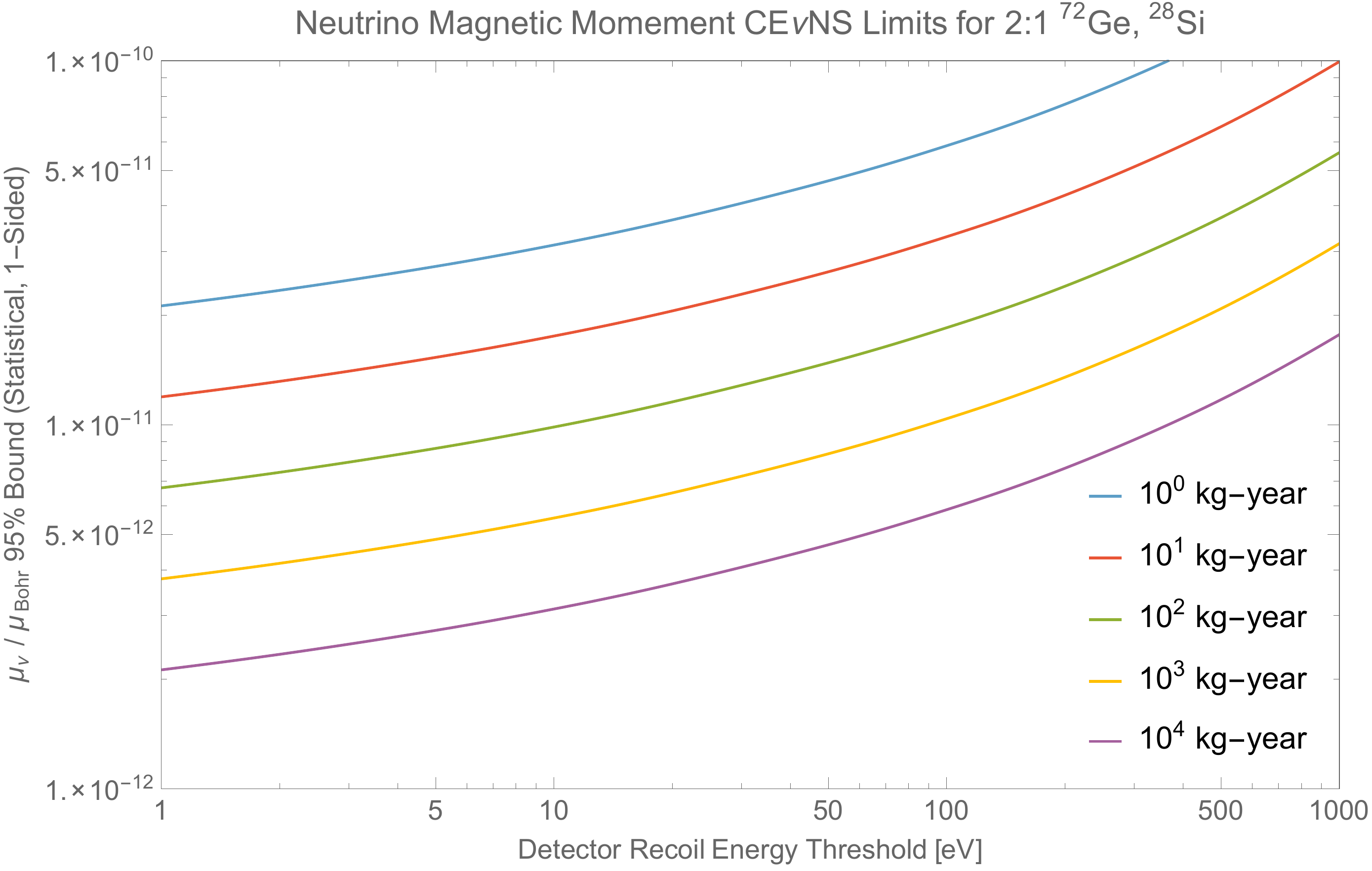}
\caption{\footnotesize
Sensitivity to the neutrino magnetic moment $\nu_\mu$ at various
scalings of the detector mass and integration time.}
\label{fig:Mu}
\end{center}
\end{figure}

In addition to the structured $Z^\prime$ contributions to the nuclear recoil rate,
it is also possible to probe for generic non-standard neutrino interaction vertices~\cite{Barranco:2005yy,Wise:2014oea},
both of the flavor-diagonal, and flavor-mixed varieties.
We adopt the notation and normalization of Ref.~\cite{Barranco:2005yy},
where, for example, $\epsilon_{ee}^{dV}$ is the coefficient for the diagonal $e/e$
neutrino current with a down quark current (vector), and  $\epsilon_{e\tau}^{uV}$
is the mixed $e/\tau$ neutrino current with an up quark.
Note that the $e/\mu$ limits are very strong from flavor
changing experiments~\cite{Davidson:2003ha}, on the order of $10^{-7}$, well beyond our
ability to resolve.  Existing $e/e$ and $e/\tau$ limits
are on the order of a few times $10^{-4}$ (cf. the second figure of Ref.~\cite{Wise:2014oea}),
converted to the Ref.~\cite{Barranco:2005yy} normalization.  The $e/e$ limits
are of the order that may be competitively probed by a low-threshold nuclear recoil detector
of the sort proposed, especially with modest extension of the target mass and/or integrated
luminosity.  Results are depicted in the Fig.~(\ref{fig:nsi})
as a function of the detector recoil threshold $T_{\rm R}^{\rm th}$
for \ger{} and \si{} in a 2:1 mass ratio, with a combined mass of 30~kg, operating
for a one-year continuous exposure, at 1~m from core.  The modular design of the
proposed detector array makes physical reconfiguration and up-scaling of the experiment
relatively straightforward.  For example, different ratios of \ger{} and \si{} may be
employed to probe relative systematics, or to accentuate enhancements in the relative
rate of a targeted BSM physics process.
Prior caveats on systematic errors, and prior elaboration of the benefits of a
differential search in both \ger{} and \si{} apply in this case as well.

It has historically been interesting to consider mutual limits on pairs of
NSI parameters, allowing for interference~\cite{Scholberg:2005qs,Barranco:2005yy}.
For example, if one considers down-type NSI coefficients with
electron-electron and electron-tau flavor mixing~\cite{Amanik:2004vm}
then the expected event rate is approximately proportional to
$[N g_V^N + (Z+2N)\,\epsilon_{ee}^{dV}]^2 + [(Z+2N)\,\epsilon_{e\tau}^{dV}]^2$.
There will be a circular ring of solutions in the $\{\epsilon_{ee}^{dV},\,\epsilon_{e\tau}^{dV}\}$
plane centered at the coordinate $\{-N/(Z+2N)\,g_V^N,0\}$, and intersecting the origin $\{0,0\}$,
for which the SM event rate is replicated by conspiracy between the offsetting NSI parameters.
The extension of this ring about its displaced center is manifestly symmetric with respect to the axes.
Modulo some thickness appropriated to systematic and statistical errors, the ring divides
discernible under-production (interior) from discernible over-production (exterior).
However, realization of this scenario requires that the NSI coefficients
are comparable in size to $g_V^N \equiv -1/2$, which is no longer
experimentally tenable~\cite{Wise:2014oea}.  Restricting to parameterizations
$\epsilon_{e[e,\tau]}^{dV} << g_V^N$ directly proximal to the origin, the
previously described circle appears instead more like a vertical line or a gently
inflecting parabola, indicating strongly preferential sensitivity to displacement
along the horizontal (flavor-diagonal) coordinate.
Moreover, there is intrinsic sensitivity to the sign of the flavor-diagonal term (with essentially
identical sensitivity to magnitude) as distinguished by over/under production,
which is not available for the off-diagonal (always over produces) coefficient.
In this context, limits for leading flavor-diagonal terms decouple
from the value of flavor-mixing NSI coefficients.

The described absence of competitive experimental bounds for the off-diagonal $e/\tau$ flavor changing
interaction is somewhat paradoxical.  In general, we emphasize that sensitivity of a nuclear recoil detector
suffers when probing very weak interactions that are expressly prohibited by SM symmetries.
This is regime where conventional experiments typically thrive because of low competing backgrounds,
but there is no straightforward mechanism in this case for discriminating the flavor structure of the 
underlying interaction vertex on an event-by-event basis.
Given sensitivity only to net deviations from an expected count of SM events,
the underlying difficulty may be recast algebraically as an absence of
rate-boosting SM cross-terms after squaring of the interaction amplitude.
For example, the leading deviation for the scenario from the prior paragraph
will come from a term proportional to $2N (Z+2N)\,g_V^N \epsilon_{ee}^{dV}$,
which is only linear in the small term $\epsilon_{ee}^{dV}$. 
Not only is the initial rate worse (like a square) for the off diagonal
coefficients, but the scaling of bounds with respect to luminosity integration
is also less steep, like $\mathcal{L}^{1/4}$ rather than $\mathcal{L}^{1/2}$.

Coherent nuclear scattering is likewise a promising channel for probing the existence
of a Majorana neutrino magnetic moment $\mu_\nu$~\cite{Kosmas:2015sqa}.
Fig.~(\ref{fig:Mu}) shows statistical search limits for the magnetic moment, using
only the leading (at low threshold) nuclear CE$\nu$NS contributions.  \ger{} and \si{} are combined in a 2:1 mass ratio.
For one unified kg-year, at a detector recoil threshold $T_{\rm R}^{\rm th} = 10$~eV,
the limit is about $3\times10^{-11}$, in units of the Bohr magneton. This is comparable 
to the present limits from terrestrial experiments~\cite{Beda:2010hk}. 
The scaling with mass and time will again be a fourth-root.  For $10^4$ kg-years, the
limit is down to about $3\times10^{-12}$, which is competitive with astrophysics sensitivity~\cite{Raffelt:1996wa}. 
As before, however, systematic errors will play a limiting role.
The event rate is proportional to just the nuclear proton count $Z$, whereas the base
CE$\nu$NS strongly integrates the count of neutrons, so that differential comparison of \ger{} and \si{}
is again very useful here to distinguish the origin of any observed event excess;
likewise, this will provide for cancellation in correlated uncertainties.

\section{Cancellation of Systematic Errors in Differential Event Rates}
\label{sec:diff} 

In this section, we extend the former presentation of absolute scale sensitivities
to new physics via \cns{} to highlight the benefits of differential
\cns{} event rate observations (cf. the approach of Ref.~\cite{Barranco:2005yy}) in multiple 
nuclei to the cancellation of persistent systematic errors.
In particular, the combination
of silicon and germanium detector elements presents the opportunity to cleanly
distinguish between various models and modes of new physics, based upon
variations in the relative coupling strength to neutrons and protons.
While intrinsically insensitive to, for example, the $Z^\prime$ mass or the
size of the neutrino magnetic moment $\mu_\nu$, this approach can reveal very clear
qualitative differences between the SM and various candidates for new physics,
in a manner that cuts through the systematic uncertainty ceiling, recovering
the potential science impact of large integrated luminosities. 

We introduce the observables
\be
\xi \quad\equiv\quad \frac{E_{\rm Ge}\,/\,B_{\rm Ge} \,-\, 1}{E_{\rm Si}\,/\,B_{\rm Si} \,-\, 1}
\quad=\quad \frac{S_{\rm Ge}\,/\,B_{\rm Ge}}{S_{\rm Si}\,/\,B_{\rm Si}}
\,,
\label{eq:gesiratio}
\ee
and
\be
\zeta \quad\equiv\quad \frac{E_{\rm Ge}}{B_{\rm Ge}} \,-\, \frac{E_{\rm Si}}{B_{\rm Si}}
\quad=\quad \frac{S_{\rm Ge}}{B_{\rm Ge}} \,-\, \frac{S_{\rm Si}}{B_{\rm Si}}
\,,
\label{eq:gesidiff}
\ee
where $E$, $B$, and $S\equiv E-B$ are the experimental total, expected standard model background,
and beyond standard model signal event counts, respectively.
Table~\ref{tab:gesiratio} itemizes signature values of the $\xi$ statistic for various $Z^\prime$ model families, and also for
nuclear scattering via the anti-neutrino magnetic moment $\mu_\nu$ coupling.  These distinctive signatures are broadly
independent of the underlying mass scale ($M_{Z^\prime}$), mixing angle ($\beta$), or coupling strength ($g'$).
The $\zeta$ statistic retains sensitivity to the new physics scale, while still allowing for the cancellation of systematic errors.
\bgroup
\def\arraystretch{1.5}
\setlength\tabcolsep{8pt}
\begin{table}[h!]
   \begin{center}
     \caption{The Eq.~(\ref{eq:gesiratio}) ratio $\xi$ of normalized BSM event
counts in \ger{} and \si{} at a detector recoil threshold $T_{\rm R}^{\rm th} = 10$~eV.}
     \label{tab:gesiratio}
     \begin{tabular}{c c c c c }
       \hline
        & $SM$ & $E_6$ & $B-L$ & $\mu_\nu$ \\ 
       \hline
       $\xi$ & $1.0$ & $0.89$ & $0.86$ & $0.43$ \\
       \hline
    \end{tabular}
  \end{center}
\end{table}
\egroup

We adopt the point of view that the theoretically calculated background counts $B$ are absolute, with zero error ($\delta B = 0$).
This is not to say, of course, that the calculation inherits no propagated uncertainty, but rather that  
differences between theory and experiment are considered to be absorbed by the experimental side.  Consequently, $\delta S = \delta E$.
Variations of the statistics in Eqs.~(\ref{eq:gesiratio},\ref{eq:gesidiff}) are then given as follows.
\be
\frac{\delta \xi}{\xi} = \frac{\delta E_{\rm Ge}}{E_{\rm Ge} - B_{\rm Ge}} - \frac{\delta E_{\rm Si}}{E_{\rm Si} - B_{\rm Si}}
\quad;\quad
\delta{\zeta} = \frac{\delta E_{\rm Ge}}{B_{\rm Ge}} - \frac{\delta E_{\rm Si}}{B_{\rm Si}}
\label{eq:diffvars}
\ee
Variations $\delta E = \delta E_{\rm Syst} + \delta E_{\rm Stat}$ will generically be composed of
both systematic and statistical components.  The systematic term is expected to be primarily correlated across detectors,
such that $\delta E_{\rm Ge}/E_{\rm Ge} \simeq \delta E_{\rm Si}/E_{\rm Si}$.  Noting that the new physics contribution
is generically expected to be small, i.e. $E \simeq B$, it is observed that systematic effects cancel to leading order
in both terms of Eq.~(\ref{eq:diffvars}), as expected.  Moreover, the residual statistical uncertainties
$\delta E_{\rm Stat} \simeq \sqrt{B}$ may be estimated as a root of the ambient SM background.  Combining these uncorrelated
statistical terms in quadrature, estimates of the standard uncertainty $\sigma$ for $\xi,\zeta$ emerge.
\be
\frac{\sigma_\xi}{\xi} = \sqrt{\frac{B_{\rm Ge}}{S^2_{\rm Ge}} + \frac{B_{\rm Si}}{S^2_{\rm Si}}}
\quad;\quad
\sigma_\zeta = \sqrt{\frac{1}{B_{\rm Ge}} + \frac{1}{B_{\rm Si}}} 
\label{eq:differrs}
\ee
The percentage error in both $\xi$ and $\zeta$ is observed to scale like $\mathcal{L}^{-1/2}$ with luminosity. 

\section{Conclusions}
\label{sec:conclusions} 
We have discussed the prospects for probing BSM physics, in particular $Z^\prime$ and non-standard 
neutrino interactions, as well as the magnetic moment of the neutrino,
using ultra-low threshold ($\sim 10$~eV) \ger{} and \si{} detectors~\cite{Mirabolfathi:2015pha}. 
This analysis is motivated by a developing experimental program in cooperation with the
Nuclear Science Center at Texas A\&M University.
We have highlighted the benefit of combining 
silicon and germanium detectors, which helps to distinguish between candidates for new physics
by leveraging distinctive couplings to neutrons vs. protons;
the benefits of this approach to bypassing the systematic error wall have also been emphasized.
The projected sensitivities to $Z^\prime$ and non-standard 
neutrino interactions are complementary to ongoing searches for new physics at the LHC,
especially with regards to the capacity for discrimination between
different models of $Z^\prime$ physics in the large statistics limit;
by extension, this specifically includes potential explanations of the diboson excess currently
reported by both CMS and ATLAS that are based upon a leptophobic $B-L$ type $Z^\prime$.
We find that the projected constraints on the neutrino magnetic moment will improve upon the terrestrial 
bound, and can become competitive with astrophysical bounds.  A summary of (statistical) bounds on leading
modes of BSM physics considered in this paper is presented in Table~\ref{tab:statistical}.
Limits for the $B-L$ model with $g' = 0.2$, as well as for the $E_6$ $\eta$-model and the SSM,
are order-comparable with the exhibited $E_6$ $\chi$-model.
Scenarios for substantially scaling up the initial mass-time exposure are tabulated.
Systematic errors are comparable to statistical errors for the baseline 
scenario, but substantially dominate in the latter cases; this
effect may be mitigated by careful application of differential event rates.
The limits for $Z^\prime$ are not greatly dependent upon the specific detector
threshold, as long as it is not too far above 100~eV.  By contrast, the magnetic
moment limits thrive with a very low-threshold detector, of the type described. 

In addition to the probes of BSM physics discussed here, ultra-low threshold detectors can be utilized
in searches for sterile neutrinos, and the detection of low energy neutrinos from the Sun. Dedicated analyses 
will be presented in forthcoming publications.

\bgroup
\def\arraystretch{1.5}
\setlength\tabcolsep{8pt}
\begin{table}[h!]
   \begin{center}
     \caption{Summary of approximate statistical sensitivity to new physics
in a 2:1 \ger{}, \si{} detector with recoil threshold $T_{\rm R}^{\rm th} = 10$~eV.
Various integrated mass-time exposures are considered, starting with near-term plans
for an order 30~kg detector operating for 1 year,
and scaling up to 1,000~kg for 5 years, and 10,000~kg for 10 years.
$Z$-prime masses are in TeV.}
     \label{tab:statistical}
     \begin{tabular}{c c c c }
       \hline
        kg-years & $M_{Z^\prime} (E_6, \chi)$ & $M_{Z^\prime} (B-L, g'=0.4)$ & $\mu_\nu/\mu_{\rm Bohr}$ \\ 
       \hline
       $30$ & $2.4$ & $4.3$ & $1.3\times10^{-11}$ \\
       $5\times10^3$ & $8.5$ & $15$ & $3.7\times10^{-12}$ \\
       $1\times10^5$ & $18$ & $32$ & $1.8\times10^{-12}$ \\
       \hline
    \end{tabular}
  \end{center}
\end{table}
\egroup

\bigskip

{\it {\bf Acknowledgements -}}
The authors acknowledge very useful discussions with Nader Mirabolfathi, Rusty Harris, and Grigory Rogachev.
BD acknowledges support from DOE Grant DE-FG02-13ER42020.
LES acknowledges support from NSF grant PHY-1522717.
JWW acknowledges support from the Mitchell Institute for Fundamental Physics and Astronomy,
NSF grant PHY-1521105, and the SHSU Department of Physics.

\bibliography{bibliography}

\begin{thebibliography}{55}
\expandafter\ifx\csname natexlab\endcsname\relax\def\natexlab#1{#1}\fi
\expandafter\ifx\csname bibnamefont\endcsname\relax
  \def\bibnamefont#1{#1}\fi
\expandafter\ifx\csname bibfnamefont\endcsname\relax
  \def\bibfnamefont#1{#1}\fi
\expandafter\ifx\csname citenamefont\endcsname\relax
  \def\citenamefont#1{#1}\fi
\expandafter\ifx\csname url\endcsname\relax
  \def\url#1{\texttt{#1}}\fi
\expandafter\ifx\csname urlprefix\endcsname\relax\def\urlprefix{URL }\fi
\providecommand{\bibinfo}[2]{#2}
\providecommand{\eprint}[2][]{\url{#2}}

\bibitem[{\citenamefont{Freedman}(1974)}]{Freedman:1973yd}
\bibinfo{author}{\bibfnamefont{D.~Z.} \bibnamefont{Freedman}},
  \bibinfo{journal}{Phys. Rev.} \textbf{\bibinfo{volume}{D9}},
  \bibinfo{pages}{1389} (\bibinfo{year}{1974}).

\bibitem[{\citenamefont{Freedman et~al.}(1977)\citenamefont{Freedman, Schramm,
  and Tubbs}}]{Freedman:1977xn}
\bibinfo{author}{\bibfnamefont{D.~Z.} \bibnamefont{Freedman}},
  \bibinfo{author}{\bibfnamefont{D.~N.} \bibnamefont{Schramm}},
  \bibnamefont{and} \bibinfo{author}{\bibfnamefont{D.~L.} \bibnamefont{Tubbs}},
  \bibinfo{journal}{Ann. Rev. Nucl. Part. Sci.} \textbf{\bibinfo{volume}{27}},
  \bibinfo{pages}{167} (\bibinfo{year}{1977}).

\bibitem[{\citenamefont{Cabrera et~al.}(1985)\citenamefont{Cabrera, Krauss, and
  Wilczek}}]{Cabrera:1984rr}
\bibinfo{author}{\bibfnamefont{B.}~\bibnamefont{Cabrera}},
  \bibinfo{author}{\bibfnamefont{L.~M.} \bibnamefont{Krauss}},
  \bibnamefont{and} \bibinfo{author}{\bibfnamefont{F.}~\bibnamefont{Wilczek}},
  \bibinfo{journal}{Phys. Rev. Lett.} \textbf{\bibinfo{volume}{55}},
  \bibinfo{pages}{25} (\bibinfo{year}{1985}).

\bibitem[{\citenamefont{Monroe and Fisher}(2007)}]{Monroe:2007xp}
\bibinfo{author}{\bibfnamefont{J.}~\bibnamefont{Monroe}} \bibnamefont{and}
  \bibinfo{author}{\bibfnamefont{P.}~\bibnamefont{Fisher}},
  \bibinfo{journal}{Phys. Rev.} \textbf{\bibinfo{volume}{D76}},
  \bibinfo{pages}{033007} (\bibinfo{year}{2007}), \eprint{0706.3019}.

\bibitem[{\citenamefont{Strigari}(2009)}]{Strigari:2009bq}
\bibinfo{author}{\bibfnamefont{L.~E.} \bibnamefont{Strigari}},
  \bibinfo{journal}{New J. Phys.} \textbf{\bibinfo{volume}{11}},
  \bibinfo{pages}{105011} (\bibinfo{year}{2009}), \eprint{0903.3630}.

\bibitem[{\citenamefont{Billard et~al.}(2014)\citenamefont{Billard, Strigari,
  and Figueroa-Feliciano}}]{Billard:2013qya}
\bibinfo{author}{\bibfnamefont{J.}~\bibnamefont{Billard}},
  \bibinfo{author}{\bibfnamefont{L.}~\bibnamefont{Strigari}}, \bibnamefont{and}
  \bibinfo{author}{\bibfnamefont{E.}~\bibnamefont{Figueroa-Feliciano}},
  \bibinfo{journal}{Phys. Rev.} \textbf{\bibinfo{volume}{D89}},
  \bibinfo{pages}{023524} (\bibinfo{year}{2014}), \eprint{1307.5458}.

\bibitem[{\citenamefont{Pospelov}(2011)}]{Pospelov:2011ha}
\bibinfo{author}{\bibfnamefont{M.}~\bibnamefont{Pospelov}},
  \bibinfo{journal}{Phys. Rev.} \textbf{\bibinfo{volume}{D84}},
  \bibinfo{pages}{085008} (\bibinfo{year}{2011}), \eprint{1103.3261}.

\bibitem[{\citenamefont{Harnik et~al.}(2012)\citenamefont{Harnik, Kopp, and
  Machado}}]{Harnik:2012ni}
\bibinfo{author}{\bibfnamefont{R.}~\bibnamefont{Harnik}},
  \bibinfo{author}{\bibfnamefont{J.}~\bibnamefont{Kopp}}, \bibnamefont{and}
  \bibinfo{author}{\bibfnamefont{P.~A.~N.} \bibnamefont{Machado}},
  \bibinfo{journal}{JCAP} \textbf{\bibinfo{volume}{1207}}, \bibinfo{pages}{026}
  (\bibinfo{year}{2012}), \eprint{1202.6073}.

\bibitem[{\citenamefont{Barranco et~al.}(2007)\citenamefont{Barranco, Miranda,
  and Rashba}}]{Barranco:2007tz}
\bibinfo{author}{\bibfnamefont{J.}~\bibnamefont{Barranco}},
  \bibinfo{author}{\bibfnamefont{O.~G.} \bibnamefont{Miranda}},
  \bibnamefont{and} \bibinfo{author}{\bibfnamefont{T.~I.}
  \bibnamefont{Rashba}}, \bibinfo{journal}{Phys. Rev.}
  \textbf{\bibinfo{volume}{D76}}, \bibinfo{pages}{073008}
  (\bibinfo{year}{2007}), \eprint{hep-ph/0702175}.

\bibitem[{\citenamefont{Adams et~al.}(2015)}]{Adams:2015ogl}
\bibinfo{author}{\bibfnamefont{C.}~\bibnamefont{Adams}} \bibnamefont{et~al.},
  in \emph{\bibinfo{booktitle}{{Workshop on the Intermediate Neutrino Program
  (WINP 2015) Upton, NY, USA, February 4-6, 2015}}} (\bibinfo{year}{2015}),
  \eprint{1503.06637},
  \urlprefix\url{http://inspirehep.net/record/1355214/files/arXiv:1503.06637.pdf}.

\bibitem[{\citenamefont{Kosmas et~al.}(2015)\citenamefont{Kosmas, Miranda,
  Papoulias, Tortola, and Valle}}]{Kosmas:2015sqa}
\bibinfo{author}{\bibfnamefont{T.~S.} \bibnamefont{Kosmas}},
  \bibinfo{author}{\bibfnamefont{O.~G.} \bibnamefont{Miranda}},
  \bibinfo{author}{\bibfnamefont{D.~K.} \bibnamefont{Papoulias}},
  \bibinfo{author}{\bibfnamefont{M.}~\bibnamefont{Tortola}}, \bibnamefont{and}
  \bibinfo{author}{\bibfnamefont{J.~W.~F.} \bibnamefont{Valle}},
  \bibinfo{journal}{Phys. Rev.} \textbf{\bibinfo{volume}{D92}},
  \bibinfo{pages}{013011} (\bibinfo{year}{2015}), \eprint{1505.03202}.

\bibitem[{\citenamefont{Mirabolfathi et~al.}(2015)\citenamefont{Mirabolfathi,
  Harris, Mahapatra, Sundqvist, Jastram, Serfass, Faiez, and
  Sadoulet}}]{Mirabolfathi:2015pha}
\bibinfo{author}{\bibfnamefont{N.}~\bibnamefont{Mirabolfathi}},
  \bibinfo{author}{\bibfnamefont{H.~R.} \bibnamefont{Harris}},
  \bibinfo{author}{\bibfnamefont{R.}~\bibnamefont{Mahapatra}},
  \bibinfo{author}{\bibfnamefont{K.}~\bibnamefont{Sundqvist}},
  \bibinfo{author}{\bibfnamefont{A.}~\bibnamefont{Jastram}},
  \bibinfo{author}{\bibfnamefont{B.}~\bibnamefont{Serfass}},
  \bibinfo{author}{\bibfnamefont{D.}~\bibnamefont{Faiez}}, \bibnamefont{and}
  \bibinfo{author}{\bibfnamefont{B.}~\bibnamefont{Sadoulet}}
  (\bibinfo{year}{2015}), \eprint{1510.00999}.

\bibitem[{\citenamefont{Wong et~al.}(2003)\citenamefont{Wong, Li, and
  Zhou}}]{Wong:2003ht}
\bibinfo{author}{\bibfnamefont{H.~T.} \bibnamefont{Wong}},
  \bibinfo{author}{\bibfnamefont{J.}~\bibnamefont{Li}}, \bibnamefont{and}
  \bibinfo{author}{\bibfnamefont{Z.~Y.} \bibnamefont{Zhou}}
  (\bibinfo{collaboration}{TEXONO}), in \emph{\bibinfo{booktitle}{{1st Yamada
  Symposium on Neutrinos and Dark Matter in Physics (YS-1 and NDM03) Nara,
  Japan, June 9-14, 2003}}} (\bibinfo{year}{2003}), \eprint{hep-ex/0307001}.

\bibitem[{\citenamefont{Goorley}(2012)}]{mcnp}
\bibinfo{author}{\bibfnamefont{e.~a.} \bibnamefont{Goorley},
  \bibfnamefont{T.}}, \bibinfo{journal}{Nucl. Tech.}
  \textbf{\bibinfo{volume}{180}}, \bibinfo{pages}{298} (\bibinfo{year}{2012}).

\bibitem[{\citenamefont{Kopeikin et~al.}(2004)\citenamefont{Kopeikin,
  Mikaelyan, and Sinev}}]{Kopeikin:2004cn}
\bibinfo{author}{\bibfnamefont{V.}~\bibnamefont{Kopeikin}},
  \bibinfo{author}{\bibfnamefont{L.}~\bibnamefont{Mikaelyan}},
  \bibnamefont{and} \bibinfo{author}{\bibfnamefont{V.}~\bibnamefont{Sinev}},
  \bibinfo{journal}{Phys. Atom. Nucl.} \textbf{\bibinfo{volume}{67}},
  \bibinfo{pages}{1892} (\bibinfo{year}{2004}), \bibinfo{note}{[Yad.
  Fiz.67,1916(2004)]}, \eprint{hep-ph/0410100}.

\bibitem[{\citenamefont{Vogel et~al.}(1981)\citenamefont{Vogel, Schenter, Mann,
  and Schenter}}]{Vogel:1980bk}
\bibinfo{author}{\bibfnamefont{P.}~\bibnamefont{Vogel}},
  \bibinfo{author}{\bibfnamefont{G.~K.} \bibnamefont{Schenter}},
  \bibinfo{author}{\bibfnamefont{F.~M.} \bibnamefont{Mann}}, \bibnamefont{and}
  \bibinfo{author}{\bibfnamefont{R.~E.} \bibnamefont{Schenter}},
  \bibinfo{journal}{Phys. Rev.} \textbf{\bibinfo{volume}{C24}},
  \bibinfo{pages}{1543} (\bibinfo{year}{1981}).

\bibitem[{\citenamefont{Kopeikin}(2012)}]{Kopeikin:2012zz}
\bibinfo{author}{\bibfnamefont{V.~I.} \bibnamefont{Kopeikin}},
  \bibinfo{journal}{Phys. Atom. Nucl.} \textbf{\bibinfo{volume}{75}},
  \bibinfo{pages}{143} (\bibinfo{year}{2012}), \bibinfo{note}{[Yad.
  Fiz.75N2,165(2012)]}.

\bibitem[{\citenamefont{Schreckenbach et~al.}(1985)\citenamefont{Schreckenbach,
  Colvin, Gelletly, and Von~Feilitzsch}}]{Schreckenbach:1985ep}
\bibinfo{author}{\bibfnamefont{K.}~\bibnamefont{Schreckenbach}},
  \bibinfo{author}{\bibfnamefont{G.}~\bibnamefont{Colvin}},
  \bibinfo{author}{\bibfnamefont{W.}~\bibnamefont{Gelletly}}, \bibnamefont{and}
  \bibinfo{author}{\bibfnamefont{F.}~\bibnamefont{Von~Feilitzsch}},
  \bibinfo{journal}{Phys. Lett.} \textbf{\bibinfo{volume}{B160}},
  \bibinfo{pages}{325} (\bibinfo{year}{1985}).

\bibitem[{\citenamefont{Venkataramaiah
  et~al.}(1985)\citenamefont{Venkataramaiah, Gopala, Basavaraju, Suryanarayana,
  and Sanjeeviah}}]{0305-4616-11-3-014}
\bibinfo{author}{\bibfnamefont{P.}~\bibnamefont{Venkataramaiah}},
  \bibinfo{author}{\bibfnamefont{K.}~\bibnamefont{Gopala}},
  \bibinfo{author}{\bibfnamefont{A.}~\bibnamefont{Basavaraju}},
  \bibinfo{author}{\bibfnamefont{S.~S.} \bibnamefont{Suryanarayana}},
  \bibnamefont{and}
  \bibinfo{author}{\bibfnamefont{H.}~\bibnamefont{Sanjeeviah}},
  \bibinfo{journal}{Journal of Physics G: Nuclear Physics}
  \textbf{\bibinfo{volume}{11}}, \bibinfo{pages}{359} (\bibinfo{year}{1985}).

\bibitem[{\citenamefont{Von~Feilitzsch
  et~al.}(1982)\citenamefont{Von~Feilitzsch, Hahn, and
  Schreckenbach}}]{VonFeilitzsch:1982jw}
\bibinfo{author}{\bibfnamefont{F.}~\bibnamefont{Von~Feilitzsch}},
  \bibinfo{author}{\bibfnamefont{A.~A.} \bibnamefont{Hahn}}, \bibnamefont{and}
  \bibinfo{author}{\bibfnamefont{K.}~\bibnamefont{Schreckenbach}},
  \bibinfo{journal}{Phys. Lett.} \textbf{\bibinfo{volume}{B118}},
  \bibinfo{pages}{162} (\bibinfo{year}{1982}).

\bibitem[{\citenamefont{Hahn et~al.}(1989)\citenamefont{Hahn, Schreckenbach,
  Colvin, Krusche, Gelletly, and Von~Feilitzsch}}]{Hahn:1989zr}
\bibinfo{author}{\bibfnamefont{A.~A.} \bibnamefont{Hahn}},
  \bibinfo{author}{\bibfnamefont{K.}~\bibnamefont{Schreckenbach}},
  \bibinfo{author}{\bibfnamefont{G.}~\bibnamefont{Colvin}},
  \bibinfo{author}{\bibfnamefont{B.}~\bibnamefont{Krusche}},
  \bibinfo{author}{\bibfnamefont{W.}~\bibnamefont{Gelletly}}, \bibnamefont{and}
  \bibinfo{author}{\bibfnamefont{F.}~\bibnamefont{Von~Feilitzsch}},
  \bibinfo{journal}{Phys. Lett.} \textbf{\bibinfo{volume}{B218}},
  \bibinfo{pages}{365} (\bibinfo{year}{1989}).

\bibitem[{\citenamefont{Mueller et~al.}(2011)}]{Mueller:2011nm}
\bibinfo{author}{\bibfnamefont{T.~A.} \bibnamefont{Mueller}}
  \bibnamefont{et~al.}, \bibinfo{journal}{Phys. Rev.}
  \textbf{\bibinfo{volume}{C83}}, \bibinfo{pages}{054615}
  (\bibinfo{year}{2011}), \eprint{1101.2663}.

\bibitem[{\citenamefont{Li and Wong}(2002)}]{Li:2001ha}
\bibinfo{author}{\bibfnamefont{H.-B.} \bibnamefont{Li}} \bibnamefont{and}
  \bibinfo{author}{\bibfnamefont{H.~T.} \bibnamefont{Wong}},
  \bibinfo{journal}{J. Phys.} \textbf{\bibinfo{volume}{G28}},
  \bibinfo{pages}{1453} (\bibinfo{year}{2002}), \eprint{hep-ex/0111002}.

\bibitem[{\citenamefont{McLaughlin}(2015)}]{McLaughlin:2015xfa}
\bibinfo{author}{\bibfnamefont{G.}~\bibnamefont{McLaughlin}},
  \bibinfo{journal}{AIP Conf. Proc.} \textbf{\bibinfo{volume}{1666}},
  \bibinfo{pages}{160001} (\bibinfo{year}{2015}).

\bibitem[{\citenamefont{Akimov et~al.}(2015)}]{Akimov:2015nza}
\bibinfo{author}{\bibfnamefont{D.}~\bibnamefont{Akimov}} \bibnamefont{et~al.}
  (\bibinfo{collaboration}{COHERENT}) (\bibinfo{year}{2015}),
  \eprint{1509.08702}.

\bibitem[{\citenamefont{Alonso et~al.}(2010)}]{Alonso:2010fs}
\bibinfo{author}{\bibfnamefont{J.}~\bibnamefont{Alonso}} \bibnamefont{et~al.}
  (\bibinfo{year}{2010}), \eprint{1006.0260}.

\bibitem[{\citenamefont{Singh and Wong}(2003)}]{Singh:2003ep}
\bibinfo{author}{\bibfnamefont{V.}~\bibnamefont{Singh}} \bibnamefont{and}
  \bibinfo{author}{\bibfnamefont{H.~T.} \bibnamefont{Wong}}
  (\bibinfo{collaboration}{TEXONO}), \bibinfo{journal}{PoS}
  \textbf{\bibinfo{volume}{AHEP2003}}, \bibinfo{pages}{053}
  (\bibinfo{year}{2003}), \eprint{nucl-ex/0412057}.

\bibitem[{\citenamefont{Wong et~al.}(2006)\citenamefont{Wong, Li, Li, Yue, and
  Zhou}}]{Wong:2005vg}
\bibinfo{author}{\bibfnamefont{H.~T.} \bibnamefont{Wong}},
  \bibinfo{author}{\bibfnamefont{H.-B.} \bibnamefont{Li}},
  \bibinfo{author}{\bibfnamefont{J.}~\bibnamefont{Li}},
  \bibinfo{author}{\bibfnamefont{Q.}~\bibnamefont{Yue}}, \bibnamefont{and}
  \bibinfo{author}{\bibfnamefont{Z.-Y.} \bibnamefont{Zhou}},
  \bibinfo{journal}{J. Phys. Conf. Ser.} \textbf{\bibinfo{volume}{39}},
  \bibinfo{pages}{266} (\bibinfo{year}{2006}).

\bibitem[{\citenamefont{Soma et~al.}(2014)}]{Soma:2014zgm}
\bibinfo{author}{\bibfnamefont{A.~K.} \bibnamefont{Soma}} \bibnamefont{et~al.}
  (\bibinfo{collaboration}{TEXONO}) (\bibinfo{year}{2014}), \eprint{1411.4802}.

\bibitem[{\citenamefont{Barbeau et~al.}(2007)\citenamefont{Barbeau, Collar, and
  Tench}}]{Barbeau:2007qi}
\bibinfo{author}{\bibfnamefont{P.~S.} \bibnamefont{Barbeau}},
  \bibinfo{author}{\bibfnamefont{J.~I.} \bibnamefont{Collar}},
  \bibnamefont{and} \bibinfo{author}{\bibfnamefont{O.}~\bibnamefont{Tench}},
  \bibinfo{journal}{JCAP} \textbf{\bibinfo{volume}{0709}}, \bibinfo{pages}{009}
  (\bibinfo{year}{2007}), \eprint{nucl-ex/0701012}.

\bibitem[{\citenamefont{Vogel and Engel}(1989)}]{Vogel:1989iv}
\bibinfo{author}{\bibfnamefont{P.}~\bibnamefont{Vogel}} \bibnamefont{and}
  \bibinfo{author}{\bibfnamefont{J.}~\bibnamefont{Engel}},
  \bibinfo{journal}{Phys. Rev.} \textbf{\bibinfo{volume}{D39}},
  \bibinfo{pages}{3378} (\bibinfo{year}{1989}).

\bibitem[{\citenamefont{Barranco et~al.}(2005)\citenamefont{Barranco, Miranda,
  and Rashba}}]{Barranco:2005yy}
\bibinfo{author}{\bibfnamefont{J.}~\bibnamefont{Barranco}},
  \bibinfo{author}{\bibfnamefont{O.~G.} \bibnamefont{Miranda}},
  \bibnamefont{and} \bibinfo{author}{\bibfnamefont{T.~I.}
  \bibnamefont{Rashba}}, \bibinfo{journal}{JHEP} \textbf{\bibinfo{volume}{12}},
  \bibinfo{pages}{021} (\bibinfo{year}{2005}), \eprint{hep-ph/0508299}.

\bibitem[{\citenamefont{Sehgal and Wanninger}(1986)}]{Sehgal:1986gn}
\bibinfo{author}{\bibfnamefont{L.~M.} \bibnamefont{Sehgal}} \bibnamefont{and}
  \bibinfo{author}{\bibfnamefont{M.}~\bibnamefont{Wanninger}},
  \bibinfo{journal}{Phys. Lett.} \textbf{\bibinfo{volume}{B171}},
  \bibinfo{pages}{107} (\bibinfo{year}{1986}).

\bibitem[{\citenamefont{Brice et~al.}(2014)}]{Brice:2013fwa}
\bibinfo{author}{\bibfnamefont{S.~J.} \bibnamefont{Brice}}
  \bibnamefont{et~al.}, \bibinfo{journal}{Phys. Rev.}
  \textbf{\bibinfo{volume}{D89}}, \bibinfo{pages}{072004}
  (\bibinfo{year}{2014}), \eprint{1311.5958}.

\bibitem[{\citenamefont{Engel}(1991)}]{Engel:1991wq}
\bibinfo{author}{\bibfnamefont{J.}~\bibnamefont{Engel}},
  \bibinfo{journal}{Phys. Lett.} \textbf{\bibinfo{volume}{B264}},
  \bibinfo{pages}{114} (\bibinfo{year}{1991}).

\bibitem[{\citenamefont{Eidelman et~al.}(2004)}]{Eidelman:2004wy}
\bibinfo{author}{\bibfnamefont{S.}~\bibnamefont{Eidelman}} \bibnamefont{et~al.}
  (\bibinfo{collaboration}{Particle Data Group}), \bibinfo{journal}{Phys.
  Lett.} \textbf{\bibinfo{volume}{B592}}, \bibinfo{pages}{1}
  (\bibinfo{year}{2004}).

\bibitem[{\citenamefont{Robinett and Rosner}(1982)}]{Robinett:1982tq}
\bibinfo{author}{\bibfnamefont{R.~W.} \bibnamefont{Robinett}} \bibnamefont{and}
  \bibinfo{author}{\bibfnamefont{J.~L.} \bibnamefont{Rosner}},
  \bibinfo{journal}{Phys. Rev.} \textbf{\bibinfo{volume}{D26}},
  \bibinfo{pages}{2396} (\bibinfo{year}{1982}).

\bibitem[{\citenamefont{Hewett and Rizzo}(1989)}]{Hewett:1988xc}
\bibinfo{author}{\bibfnamefont{J.~L.} \bibnamefont{Hewett}} \bibnamefont{and}
  \bibinfo{author}{\bibfnamefont{T.~G.} \bibnamefont{Rizzo}},
  \bibinfo{journal}{Phys. Rept.} \textbf{\bibinfo{volume}{183}},
  \bibinfo{pages}{193} (\bibinfo{year}{1989}).

\bibitem[{\citenamefont{Carena et~al.}(2004)\citenamefont{Carena, Daleo,
  Dobrescu, and Tait}}]{Carena:2004xs}
\bibinfo{author}{\bibfnamefont{M.}~\bibnamefont{Carena}},
  \bibinfo{author}{\bibfnamefont{A.}~\bibnamefont{Daleo}},
  \bibinfo{author}{\bibfnamefont{B.~A.} \bibnamefont{Dobrescu}},
  \bibnamefont{and} \bibinfo{author}{\bibfnamefont{T.~M.~P.}
  \bibnamefont{Tait}}, \bibinfo{journal}{Phys. Rev.}
  \textbf{\bibinfo{volume}{D70}}, \bibinfo{pages}{093009}
  (\bibinfo{year}{2004}), \eprint{hep-ph/0408098}.

\bibitem[{\citenamefont{Langacker}(2009)}]{Langacker:2008yv}
\bibinfo{author}{\bibfnamefont{P.}~\bibnamefont{Langacker}},
  \bibinfo{journal}{Rev. Mod. Phys.} \textbf{\bibinfo{volume}{81}},
  \bibinfo{pages}{1199} (\bibinfo{year}{2009}), \eprint{0801.1345}.

\bibitem[{\citenamefont{Laha et~al.}(2014)\citenamefont{Laha, Dasgupta, and
  Beacom}}]{Laha:2013xua}
\bibinfo{author}{\bibfnamefont{R.}~\bibnamefont{Laha}},
  \bibinfo{author}{\bibfnamefont{B.}~\bibnamefont{Dasgupta}}, \bibnamefont{and}
  \bibinfo{author}{\bibfnamefont{J.~F.} \bibnamefont{Beacom}},
  \bibinfo{journal}{Phys. Rev.} \textbf{\bibinfo{volume}{D89}},
  \bibinfo{pages}{093025} (\bibinfo{year}{2014}), \eprint{1304.3460}.

\bibitem[{\citenamefont{Ng and Beacom}(2014)}]{Ng:2014pca}
\bibinfo{author}{\bibfnamefont{K.~C.~Y.} \bibnamefont{Ng}} \bibnamefont{and}
  \bibinfo{author}{\bibfnamefont{J.~F.} \bibnamefont{Beacom}},
  \bibinfo{journal}{Phys. Rev.} \textbf{\bibinfo{volume}{D90}},
  \bibinfo{pages}{065035} (\bibinfo{year}{2014}), \bibinfo{note}{[Erratum:
  Phys. Rev.D90,no.8,089904(2014)]}, \eprint{1404.2288}.

\bibitem[{\citenamefont{Basso et~al.}(2009)\citenamefont{Basso, Belyaev,
  Moretti, Pruna, and Shepherd-Themistocleous}}]{Basso:2009gg}
\bibinfo{author}{\bibfnamefont{L.}~\bibnamefont{Basso}},
  \bibinfo{author}{\bibfnamefont{A.}~\bibnamefont{Belyaev}},
  \bibinfo{author}{\bibfnamefont{S.}~\bibnamefont{Moretti}},
  \bibinfo{author}{\bibfnamefont{G.~M.} \bibnamefont{Pruna}}, \bibnamefont{and}
  \bibinfo{author}{\bibfnamefont{C.~H.} \bibnamefont{Shepherd-Themistocleous}},
  \bibinfo{journal}{PoS} \textbf{\bibinfo{volume}{EPS-HEP2009}},
  \bibinfo{pages}{242} (\bibinfo{year}{2009}), \eprint{0909.3113}.

\bibitem[{\citenamefont{Basso et~al.}(2011)\citenamefont{Basso, Belyaev,
  Moretti, Pruna, and Shepherd-Themistocleous}}]{Basso:2010pe}
\bibinfo{author}{\bibfnamefont{L.}~\bibnamefont{Basso}},
  \bibinfo{author}{\bibfnamefont{A.}~\bibnamefont{Belyaev}},
  \bibinfo{author}{\bibfnamefont{S.}~\bibnamefont{Moretti}},
  \bibinfo{author}{\bibfnamefont{G.~M.} \bibnamefont{Pruna}}, \bibnamefont{and}
  \bibinfo{author}{\bibfnamefont{C.~H.} \bibnamefont{Shepherd-Themistocleous}},
  \bibinfo{journal}{Eur. Phys. J.} \textbf{\bibinfo{volume}{C71}},
  \bibinfo{pages}{1613} (\bibinfo{year}{2011}), \eprint{1002.3586}.

\bibitem[{\citenamefont{Gershtein et~al.}(2013)}]{Gershtein:2013iqa}
\bibinfo{author}{\bibfnamefont{Y.}~\bibnamefont{Gershtein}}
  \bibnamefont{et~al.}, in \emph{\bibinfo{booktitle}{{Community Summer Study
  2013: Snowmass on the Mississippi (CSS2013) Minneapolis, MN, USA, July
  29-August 6, 2013}}} (\bibinfo{year}{2013}), \eprint{1311.0299},
  \urlprefix\url{http://inspirehep.net/record/1263192/files/arXiv:1311.0299.pdf}.

\bibitem[{\citenamefont{Cao}(2012)}]{Cao:2011gb}
\bibinfo{author}{\bibfnamefont{J.}~\bibnamefont{Cao}}, \bibinfo{journal}{Nucl.
  Phys. Proc. Suppl.} \textbf{\bibinfo{volume}{229-232}}, \bibinfo{pages}{205}
  (\bibinfo{year}{2012}), \eprint{1101.2266}.

\bibitem[{\citenamefont{Aad et~al.}(2015)}]{Aad:2015owa}
\bibinfo{author}{\bibfnamefont{G.}~\bibnamefont{Aad}} \bibnamefont{et~al.}
  (\bibinfo{collaboration}{ATLAS}) (\bibinfo{year}{2015}), \eprint{1506.00962}.

\bibitem[{\citenamefont{Khachatryan et~al.}(2014)}]{Khachatryan:2014hpa}
\bibinfo{author}{\bibfnamefont{V.}~\bibnamefont{Khachatryan}}
  \bibnamefont{et~al.} (\bibinfo{collaboration}{CMS}), \bibinfo{journal}{JHEP}
  \textbf{\bibinfo{volume}{08}}, \bibinfo{pages}{173} (\bibinfo{year}{2014}),
  \eprint{1405.1994}.

\bibitem[{\citenamefont{Gao et~al.}(2015)\citenamefont{Gao, Ghosh, Sinha, and
  Yu}}]{Gao:2015irw}
\bibinfo{author}{\bibfnamefont{Y.}~\bibnamefont{Gao}},
  \bibinfo{author}{\bibfnamefont{T.}~\bibnamefont{Ghosh}},
  \bibinfo{author}{\bibfnamefont{K.}~\bibnamefont{Sinha}}, \bibnamefont{and}
  \bibinfo{author}{\bibfnamefont{J.-H.} \bibnamefont{Yu}}
  (\bibinfo{year}{2015}), \eprint{1506.07511}.

\bibitem[{\citenamefont{Wise and Zhang}(2014)}]{Wise:2014oea}
\bibinfo{author}{\bibfnamefont{M.~B.} \bibnamefont{Wise}} \bibnamefont{and}
  \bibinfo{author}{\bibfnamefont{Y.}~\bibnamefont{Zhang}},
  \bibinfo{journal}{Phys. Rev.} \textbf{\bibinfo{volume}{D90}},
  \bibinfo{pages}{053005} (\bibinfo{year}{2014}), \eprint{1404.4663}.

\bibitem[{\citenamefont{Davidson et~al.}(2003)\citenamefont{Davidson,
  Pena-Garay, Rius, and Santamaria}}]{Davidson:2003ha}
\bibinfo{author}{\bibfnamefont{S.}~\bibnamefont{Davidson}},
  \bibinfo{author}{\bibfnamefont{C.}~\bibnamefont{Pena-Garay}},
  \bibinfo{author}{\bibfnamefont{N.}~\bibnamefont{Rius}}, \bibnamefont{and}
  \bibinfo{author}{\bibfnamefont{A.}~\bibnamefont{Santamaria}},
  \bibinfo{journal}{JHEP} \textbf{\bibinfo{volume}{03}}, \bibinfo{pages}{011}
  (\bibinfo{year}{2003}), \eprint{hep-ph/0302093}.

\bibitem[{\citenamefont{Scholberg}(2006)}]{Scholberg:2005qs}
\bibinfo{author}{\bibfnamefont{K.}~\bibnamefont{Scholberg}},
  \bibinfo{journal}{Phys. Rev.} \textbf{\bibinfo{volume}{D73}},
  \bibinfo{pages}{033005} (\bibinfo{year}{2006}), \eprint{hep-ex/0511042}.

\bibitem[{\citenamefont{Amanik et~al.}(2005)\citenamefont{Amanik, Fuller, and
  Grinstein}}]{Amanik:2004vm}
\bibinfo{author}{\bibfnamefont{P.~S.} \bibnamefont{Amanik}},
  \bibinfo{author}{\bibfnamefont{G.~M.} \bibnamefont{Fuller}},
  \bibnamefont{and}
  \bibinfo{author}{\bibfnamefont{B.}~\bibnamefont{Grinstein}},
  \bibinfo{journal}{Astropart. Phys.} \textbf{\bibinfo{volume}{24}},
  \bibinfo{pages}{160} (\bibinfo{year}{2005}), \eprint{hep-ph/0407130}.

\bibitem[{\citenamefont{Beda et~al.}(2010)\citenamefont{Beda, Brudanin, Egorov,
  Medvedev, Pogosov, Shirchenko, and Starostin}}]{Beda:2010hk}
\bibinfo{author}{\bibfnamefont{A.~G.} \bibnamefont{Beda}},
  \bibinfo{author}{\bibfnamefont{V.~B.} \bibnamefont{Brudanin}},
  \bibinfo{author}{\bibfnamefont{V.~G.} \bibnamefont{Egorov}},
  \bibinfo{author}{\bibfnamefont{D.~V.} \bibnamefont{Medvedev}},
  \bibinfo{author}{\bibfnamefont{V.~S.} \bibnamefont{Pogosov}},
  \bibinfo{author}{\bibfnamefont{M.~V.} \bibnamefont{Shirchenko}},
  \bibnamefont{and} \bibinfo{author}{\bibfnamefont{A.~S.}
  \bibnamefont{Starostin}} (\bibinfo{year}{2010}), \eprint{1005.2736}.

\bibitem[{\citenamefont{Raffelt}(1996)}]{Raffelt:1996wa}
\bibinfo{author}{\bibfnamefont{G.~G.} \bibnamefont{Raffelt}},
  \emph{\bibinfo{title}{{Stars as laboratories for fundamental physics}}}
  (\bibinfo{year}{1996}), ISBN \bibinfo{isbn}{9780226702728},
  \urlprefix\url{http://wwwth.mpp.mpg.de/members/raffelt/mypapers/199613.pdf}.

\end{thebibliography}

\end{document}